\documentclass[twocolumn]{svjour3}    
\smartqed 
\makeatletter
\expandafter\let\csname opt@amsmath.sty\endcsname\relax
\AtBeginDocument{
	\mathindent=15pt 
	\@mathmargin\@centering} 
\makeatother
\usepackage{graphicx}
\usepackage{caption}
\usepackage{cite}
\usepackage{amsmath,amssymb,amsfonts}
\usepackage{algorithmic}
\usepackage{textcomp}
\usepackage{xcolor}
\usepackage{multirow}
\usepackage{adjustbox}
\usepackage{booktabs}
\usepackage{enumitem}
\usepackage{array}
\usepackage{tabularx}
\usepackage{hyperref}
\usepackage{subcaption}

\usepackage{soul}
\usepackage[linesnumbered,ruled,vlined]{algorithm2e}
\usepackage{boldline}
\usepackage{float}
\newcommand{\kw}[1]{\text{{#1}}\xspace}
\newcommand{\tkskq}{\kw{TkQ}}
\newcommand{\tkskqs}{\kw{TkQs}}
\newcommand{\Framework}{\kw{LIST}}

\newcommand{\revision}[1]{\textcolor{black}{#1}}

\begin{document}

\title{LIST: Learning to Index Spatio-Textual Data for 
Embedding based Spatial Keyword Queries}
\author{Ziqi Yin$\scriptsize^{\dag}$         \and
        Shanshan Feng$\scriptsize^{\ddag}$ \and
        Shang Liu$\scriptsize^{\dag}$ \and
        Gao Cong$\scriptsize^{\dag}$ \and
        Yew Soon Ong$\scriptsize^{\dag}$ \and
        Bin Cui$^*$
}

\institute{Z. Yin \at
        \email{ziqi003@e.ntu.edu.sg}           
        \and
        S. Feng  \at
        \email{victor\_fengss@foxmail.com} 
        \and
        S. Liu \at
        \email{shang006@e.ntu.edu.sg}         
        \and
        G. Cong \at
        \email{gaocong@e.ntu.edu.sg}         
        \and
        Y. Ong \at
        \email{asysong@e.ntu.edu.sg}    
        \and
        B. Cui \at
        \email{bin.cui@pku.edu.cn} \at
	\begin{description}
		\item[$\scriptsize^{\dag}$] 
            College of Computing and Data Science, Nanyang Technological University, Singapore
		\item[$\scriptsize^{\ddag}$] Centre for Frontier AI Research, A*STAR, Singapore
		\item[$^*$] School of Computer Science, Peking University, China
	\end{description}
}

\date{Received: date / Accepted: date}

\maketitle

\begin{abstract}
With the proliferation of spatio-textual data, Top-k KNN spatial keyword queries (\tkskqs), which return a list of objects based on a ranking function that considers both spatial and textual relevance, have found many real-life applications. 
To efficiently handle \tkskqs, many indexes have been developed, but the effectiveness of \tkskq is limited. 
To improve effectiveness, several deep learning models have recently been proposed, but they suffer severe efficiency issues and there are no efficient indexes specifically designed to accelerate the top-k search process for these deep learning models.


To tackle these issues, \revision{we consider embedding based spatial keyword queries, which capture the semantic meaning of query keywords and object descriptions in two separate embeddings to evaluate textual relevance. Although various 
models can be used to generate these embeddings, no indexes have been specifically designed for such queries. To fill this gap, we propose \Framework, a novel machine learning based Approximate Nearest Neighbor Search index that \underline{L}earns to \underline{I}ndex the \underline{S}patio-\underline{T}extual data. 
\Framework utilizes a new learning-to-cluster technique to group relevant queries and objects together while separating irrelevant queries and objects. There are two key challenges in building an effective and efficient index, i.e., the absence of high-quality labels and the unbalanced clustering results. We develop a novel pseudo-label generation technique to address the two challenges. Additionally, we introduce a learning based spatial relevance model that can integrates with various text relevance models to form a lightweight yet effective relevance for reranking objects retrieved by \Framework. Experimental results show that (1) our lightweight embedding based relevance model significantly outperforms  state-of-the-art relevance models; (2) \Framework outperforms state-of-the-art indexes, providing a better trade-off between effectiveness and efficiency.}

\end{abstract}


\section{Introduction}
\label{sec:intro}
With the proliferation of mobile Internet, spatio-textual (a.k.a geo-textual) data is being increasingly generated. Examples of spatio-textual data include (1) web pages with geographical information; (2) user-generated text content with location information, such as geo-tagged tweets and reviews related to local stores; (3) Points of Interest (POI) in local business websites or location-based apps~\cite{DBLP:journals/pvldb/ChenLCLBLGZ21}; (4) multimedia data that  contains both text and geographical location, like photos shared on social platforms that provide both textual descriptions and geographic location. Meanwhile, with the prevalence of smartphones, accessing and querying spatio-textual data has become increasingly frequent. This trend calls for techniques to process spatial keyword queries efficiently and effectively, which take query keywords and location as input and return objects that match the given requirements. \revision{An example query is to search for a `delicious pizza restaurant' that is close to the user’s location. A returned object could be a nearby restaurant named `Pizza Palace'.}

Spatial keyword queries have applications in many real-world scenarios such as geographic search engines~\cite{DBLP:conf/sigmod/ChenSM06}, location-based services~\cite{DBLP:conf/cikm/ZhouXWGM05}, and local web advertising tailored to specific regions~\cite{dong2021continuous}. To meet diverse user needs, various types of spatial keyword queries have been introduced~\cite{congEfficientRetrievalTopk2009,DBLP:conf/sigmod/ChenSM06,cary2010efficient,DBLP:journals/vldb/ChenCCJ21}. Among them, the \underline{T}op-\underline{k} KNN Spatial Keyword \underline{Q}uery (\tkskq)~\cite{congEfficientRetrievalTopk2009} retrieves the top-k geo-textual objects according to a ranking function that considers both textual and spatial relevance. Specifically, \tkskq computes textual relevance by traditional information retrieval models such as BM25 and TF-IDF~\cite{DBLP:journals/ftir/RobertsonZ09,DBLP:journals/tois/WuLWK08} and uses a linear function of distance between query location and object location to evaluate spatial relevance (as to be formulated in Section~\ref{sec:preliminary}). According to the experimental evaluation~\cite{liu2023effectiveness}, \tkskqs return more relevant objects compared to several other spatial keyword queries like the Boolean KNN Query~\cite{cary2010efficient}. Most of the existing studies~\cite{congEfficientRetrievalTopk2009,DBLP:conf/icde/FelipeHR08,DBLP:journals/tkde/LiLZLLW11, DBLP:conf/ssd/RochaGJN11} on spatial keyword queries focus on improving the efficiency of handling spatial keyword queries. 
As such, various indexes~\cite{DBLP:journals/vldb/ChenCCJ21} and corresponding query processing algorithms have been developed. 

\noindent\textbf{Motivations.} Despite various indexes~\cite{DBLP:journals/vldb/ChenCCJ21} have been developed to expedite the top-k search process of \tkskqs, the effectiveness of \tkskqs is limited. As discussed earlier, \tkskq uses traditional models such as BM25 to compute textual relevance, but these models rely on exact word matching to evaluate textual relevance and thus suffer from the word mismatch issue~\cite{DBLP:conf/emnlp/KarpukhinOMLWEC20,renRocketQAv2JointTraining2021,DBLP:conf/naacl/QuDLLRZDWW21}, e.g., synonyms that consist of different tokens may convey the same or similar meanings, which limits their effectiveness (as to be detailed in Section~\ref{sec:motivations}). \revision{For example, consider searching for an `Italian restaurant' on a location-based app like Foursquare. Even if a nearby restaurant named `Pasta House' exists, traditional models 
will not be able to retrieve this result because they cannot semantically match `Italian restaurant' with `Pasta House', it will receive 
zero textual relevance score.}




To improve the effectiveness of \tkskq, several deep learning based relevance models~\cite{liu2023effectiveness,DBLP:conf/sigir/DingCXHLZX23,DBLP:conf/aaai/ZhaoPWCYZMCYQ19} have recently been proposed, but they suffer from severe efficiency issue. \revision{For instance, DrW~\cite{liu2023effectiveness} employs the BERT model~\cite{devlin2018bert} to generate word embeddings and identifies top-k relevant terms from object description for each query keyword based on the word embeddings. It then uses an attention mechanism to aggregate the relevance scores between each keyword and its top-k relevant terms, determining the overall textual relevance. In our experiment, DrW takes over 7 seconds to answer a query in the Geo-Glue dataset using brute-force search. Moreover, to the best of our knowledge, there are no efficient indexes 
designed to expedite the top-k search process for these deep relevance models. 
}

\noindent\revision{\textbf{Objective and Challenges.} To this end, we aim to develop an efficient index for deep learning models to answer spatial keyword queries. However, these models, such as DrW, often suffer from high latency and it is difficult to design indexes for them due to their complex word interaction functions, which rely on word embeddings to calculate the word-level similarity between query keywords and object descriptions for evaluating textual relevance. According to~\cite{guo2020deep}, these functions not only result in high latency but also cannot be pre-calculated until the query-object text pairs are seen, making it challenging to build an index for them.}

\revision{To address this, we consider the embedding based spatial keyword queries, which capture the semantic meanings of query keywords and object descriptions into two separate embeddings and evaluate textual relevance based on the two embeddings. 
Many 
text 
models can be used for the embedding based spatial keyword queries, ranging from earlier Word2Vec~\cite{mikolov2013efficient} 
to the recent advancements in pre-trained language models~\cite{zhao2024dense}.
} 

\revision{The challenge lies in developing an index for embedding based spatial keyword queries.}
Existing geo-textual indexes are developed based on traditional models such as BM25 and TF-IDF, which 
cannot be used to handle embedding based spatial keyword queries. Although Approximate Nearest Neighbor Search (ANNS) indexes~\cite{qin2020similarity,DBLP:journals/pami/WangZSSS18,DBLP:journals/pvldb/WangXY021} are designed for embedding retrieval, these ANNS indexes do not consider 
the spatial factor, which is essential for spatial keyword queries. Directly using these ANNS indexes for embedding based spatial keyword queries results in severe degradation of effectiveness (as to be shown in Section~\ref{sec:exp-effectiveness}). An intuitive idea is to adapt these ANNS indexes to incorporate the spatial factor. For instance, the IVF index~\cite{jegouProductQuantizationNearest2011} clusters embeddings using the K-means algorithm and route queries to a small number of close clusters to reduce the search space. However, incorporating spatial factors into the K-means algorithm is challenging as it is hard to set the weight to balance spatial and embedding similarities during index construction. Manually setting the weights is not only laborious but also ineffective, resulting in inferior effectiveness (as to be shown in Section~\ref{sec:exp-effectiveness}). It is still an open problem to design an effective and efficient ANNS index to support embedding based spatial keyword queries.

To fill this gap, we develop a new machine learning based ANNS index \revision{that is applicable to any relevance models for embedding based spatial keyword queries only if query and objects are represented by two embeddings.} To 
cluster both spatial and textual relevant objects together without manually setting the weight between spatial relevance and textual relevance, our proposed index utilizes the learning-to-cluster technique, which was originally developed for image clustering, to learn from pairwise relevant and irrelevant query-object pairs, thereby clustering relevant objects and queries together while separating the irrelevant ones. Existing learning-to-cluster studies~\cite{Hsu16_KCL,Hsu18_L2C} demonstrate that high-quality pairwise similar/dis-similar labels are essential for training. Although these pairwise labels are easy to obtain for images,
such pairwise positive labels between 
queries and 
objects in our problem are very sparse
, and high-quality negative labels are absent. Additionally, when the number of clusters is large, the existing learning-to-cluster technique will produce a highly skewed cluster distribution~\cite{Hsu16_KCL}, which will hurt the index's efficiency if the index is built based on the cluster results. To address these challenges, we propose a novel pseudo-negative label generation method, which employs the trained relevance model to produce high-quality pseudo-negative labels. 
Through these informative labels, our index learns to precisely cluster relevant queries and objects together while separating the irrelevant ones, thereby producing a precise and balanced clustering result and constructing an effective and efficient ANNS index.

\revision{We call the proposed index as \Framework, which \underline{l}earns to \underline{i}ndex \underline{s}patio-\underline{t}extual data for answering embedding based spatial keyword queries. \Framework employs a relevance model to pre-compute embeddings for geo-textual objects, and learns to index these embeddings. 
Given an embedding-based spatial keyword query, \Framework routes the query to a subset of clusters to reduce the search space. Subsequently, the relevance model is applied to re-rank the retrieved objects within the relevant clusters (as to be detailed in Section~\ref{sec:LIST}).}

\revision{In addition, we improve the spatial relevance for embedding based spatial keyword queries. Existing relevance models such as DrW typically 
use the linear function of distance, 
assuming that spatial preference decreases linearly with distance. 
However, this assumption was not examined previously. We find it does not hold on our real-life datasets, and that  the spatial preference 
exhibits a significant (non-linear) decrease with the increase of distance (as to be detailed in Section~\ref{sec:motivations}). This motivates us to design a new spatial relevance model to fit this real-world pattern better. A straightforward approach would be to employ an exponential function of distance. However, our experiments indicate that this simple method is even less effective compared to the linear function (as to be shown in Section~\ref{sec:exp-ablation}).} 
\revision{Instead, we design a learning based spatial relevance model, which learns from real-world datasets to better evaluate spatial relevance. The spatial model is built on two real-world patterns: (1) spatial relevance increases as distance decreases, and (2) spatial relevance increases in a stepwise manner rather than continuously as distance decreases. Therefore, we designed the model as a learnable monotonic step function. Additionally, we introduce a weight learning module to adaptively learn a weight for embedding based spatial keyword queries to balance the textual and spatial relevance. 
}

The contributions of this work are summarized as follows:
\begin{itemize}[leftmargin=*,topsep=0pt]
\item \textbf{New Index.} We develop a novel machine learning based ANNS index that is tailored for embedding based spatial keyword queries and accelerating the top-k search process of the proposed deep relevance model. Different from existing ANNS indexes, it utilizes the learning-to-cluster technique to cluster relevant objects and queries together while separating the irrelevant ones. To build an effective and efficient index, we propose a novel pseudo-label generation approach. To the best of our knowledge, this is the first index designed for embedding based spatial keyword queries and deep relevance model. Additionally, this is the first geo-textual index that employs neural networks for retrieval without relying on an explicit tree structure.

\item \revision{\textbf{New Spatial Relevance Model.} We develop a novel learning based spatial relevance module, which is capable of learning from real-world datasets to better evaluate the spatial relevance. This model can integrate with various 
textual relevance models 
for embedding based spatial keyword queries.}


\item \revision{\textbf{Extensive Experiments.} We extensively evaluate the effectiveness and efficiency of our solution \Framework
on three real-world datasets. Experimental results show that (1) our lightweight embedding based relevance model
significantly outperforms the state-of-the-art relevance models for effectiveness by an improvement up to 31.60\% and 59.92\% in terms of NDCG@1 and Recall@10, respectively; 
(2) \Framework outperforms existing state-of-the-art indexes on the three datasets, providing a better trade-off between effectiveness and efficiency.} 


\end{itemize}

\section{Related Work}
\label{sec:related}

\noindent \textbf{Spatio-Textual Relevance Models.}
Spatial keyword queries have attracted extensive attention and many types of spatial keyword queries~\cite{congEfficientRetrievalTopk2009,DBLP:conf/sigmod/ChenSM06,cary2010efficient} have been proposed. Among them, Top-k KNN Spatial Keyword Query (\tkskq)~\cite{congEfficientRetrievalTopk2009} aims to retrieve top-k geo-textual objects based on a ranking function that evaluates both spatial and textual relevance.
Specifically, \tkskq computes the textual relevance with traditional retrieval models like BM25 and TF-IDF~\cite{DBLP:journals/ftir/RobertsonZ09,DBLP:journals/tois/WuLWK08} and exploits a simple linear function of distance for spatial relevance. However, these traditional methods have limited effectiveness.

To enhance the effectiveness, several deep learning based methods~\cite{DBLP:conf/aaai/ZhaoPWCYZMCYQ19,liu2023effectiveness,DBLP:conf/sigir/DingCXHLZX23}
have been developed for query-POI matching, which is essentially spatial keyword query. PALM~\cite{DBLP:conf/aaai/ZhaoPWCYZMCYQ19} considers geographic information by using location embedding techniques and combines it with textual word semantic representations for query-POI matching. 
DrW~\cite{liu2023effectiveness} computes the deep textual relevance on the term level of query keywords and object descriptions. It uses the attention mechanism to aggregate the scores of each term for spatial keyword queries and design a learning-based method to learn a query-dependent weight to balance textual and spatial relevance. MGeo~\cite{DBLP:conf/sigir/DingCXHLZX23} employs a geographic encoder and a multi-modal interaction module, treating geographic context as a new modality and using text information as another modality. MGeo aligns these two modalities into the same latent space and computes relevance scores based on the produced representations. 
However, these deep learning based methods focus on improving the effectiveness but ignore the efficiency issue. 

Different from previous 
deep learning based relevance models that rely on word embeddings and complex models to compute textual relevance, 
\revision{
we consider embedding based spatial keyword queries, which transform query keywords and object descriptions into two separate embeddings and computes textual relevance based on the two embeddings. We also propose a new learning based spatial module to learn from real-world dataset to better estimate spatial relevance.}

\noindent \textbf{Spatio-Textual Indexes.} 
Various spatio-textual indexes~\cite{DBLP:conf/sigmod/ChenSM06,cary2010efficient,DBLP:conf/icde/FelipeHR08,congEfficientRetrievalTopk2009, DBLP:journals/tkde/LiLZLLW11, DBLP:conf/ssd/RochaGJN11, DBLP:conf/icde/ZhangZZL13,tao2013fast,vaid2005spatio,DBLP:conf/cikm/GobelHNB09,DBLP:conf/cikm/ZhouXWGM05,DBLP:conf/sigmod/LuLC11,DBLP:conf/edbt/ZhangTT13,DBLP:conf/icde/ZhangOT10,DBLP:conf/icde/ZhangCMTK09, DBLP:journals/pacmmod/Sheng0F00C023,DBLP:conf/gis/Liu022} have been designed to efficiently answer spatial keyword queries. However, they are all designed for traditional retrieval models (e.g., TF-IDF), and 
are unsuitable for accelerating the top-k search process of these deep learning based methods. 

Several indexes~\cite{DBLP:journals/www/QianXZZZ18,DBLP:journals/geoinformatica/ChenXZZLFZ20} have been introduced to incorporate semantic representations into the \tkskq scheme. 
For example, S$^2$R-Tree~\cite{DBLP:journals/geoinformatica/ChenXZZLFZ20} projects the word embeddings to
an $m$-dimensional vector using a pivot-based technique ($m$ as low as 2). Consequently, it employs R-trees to index objects based on their geo-locations and $m$-dimensional vectors hierarchically. When the dimensionality reaches hundreds, which are common for embeddings, such methods are no better than a linear scan 
due to the curse of dimensionality~\cite{indyk1998approximate}. 

Additionally, there exists a learned geo-textual index called WISK~\cite{DBLP:journals/pacmmod/Sheng0F00C023}, 
which utilizes query workloads to partition
geo-textual data to build a tree-based index and employ reinforcement learning techniques to optimize the index. However, WISK is designed for spatial keyword range queries, which treats query keywords and query region as Boolean filters, only retrieves objects containing all the query keywords within the given query region, and is not designed to support other queries like the \tkskq, which uses keywords to compute textual relevance.

Different from previous spatio-textual indexes developed to expedite the top-k search for a ranking function that uses traditional relevance models 
like BM25 and TF-IDF, which depend on exact word matching to compute textual relevance, our goal is to design an index that accelerates the top-k search for a deep learning based relevance model \revision{that is developed for embedding based spatial keyword queries.} 
It is still an open problem for geo-textual data.

\noindent \textbf{Deep Textual Relevance Models and Approximate Nearest Neighbor Search Indexes.}
%
Our work is related to deep textual relevance models and the corresponding Approximate Nearest Neighbor Search (ANNS) index techniques. 
The deep textual relevance models can be broadly classified into two categories: interaction-focused models and representation-focused models~\cite{guo2020deep}. The first category of models (e.g., ARC-II~\cite{DBLP:conf/nips/HuLLC14} and MatchPyramid~\cite{DBLP:conf/aaai/PangLGXWC16}) calculates the word-level similarity between queries and documents for textual relevance. DrW~\cite{liu2023effectiveness} belongs to this category. Although this category of methods may have better effectiveness, these methods are usually computationally expensive. The second category of models (e.g., DSSM~\cite{huang2013learning}) extracts global semantic representation for input text and uses functions like the inner product to compute the relevance score between representations. This category of models used to be less effective than the first category of models. However, with the emergence of 
Pre-trained Large Language Models (PLMs), which extract global semantic representation from textual content, it has become a well-established paradigm in document retrieval~\cite{DBLP:journals/tois/GuoCFSZC22,DBLP:journals/corr/abs-2211-14876,ECML_23_zhou2023master_dense_retrieval,SIGIR_23_wen2023offline_dense_retrieval,DBLP:conf/naacl/QuDLLRZDWW21}. For the second category of methods, to support efficient online retrieval, the learned sentence embeddings are usually pre-computed offline and indexed by the ANNS indexes~\cite{guo2020deep}. Note that, models in the first category like DrW~\cite{liu2023effectiveness} are not efficient for online computation and retrieval 
since the complex word interaction function cannot be pre-calculated until we see the input query-object pairs according to~\cite{guo2020deep}.

ANNS indexes are developed to expedite the top-k search on high-dimensional embeddings.
These techniques can be broadly categorized into two types. The first type focuses on searching a subset of the database. The representative methods include inverted file index (IVF) based methods~\cite{DBLP:journals/pami/BabenkoL15,jegouProductQuantizationNearest2011,wang2020deltapq}, hashing-based methods~\cite{luoSurveyDeepHashing2023, liu2014sk, liu2019lsh,zheng2020pm}, and graph-based methods~\cite{malkovEfficientRobustApproximate2020, DBLP:journals/pvldb/WangXY021}. These ANNS indexes typically rely on heuristic algorithms to find a subset of candidates. Specifically, Inverted File Index (IVF)~\cite{jegouProductQuantizationNearest2011} partitions data into clusters using the k-means algorithm and routes queries to a subset of close clusters based on the query's distance to the clusters' centroids. Graph-based algorithms like Hierarchical Navigable Small World (HNSW) index~\cite{malkovEfficientRobustApproximate2020} construct proximity graphs and perform a beam search on the graph for a given query to find similar embeddings. Hashing-based methods like Locality-Sensitive Hashing (LSH) index~\cite{DBLP:journals/corr/WangSSJ14} generate top-k candidates by hashing vectors into buckets and then retrieving the closest items from those buckets as candidates. The other type aims to accelerate the search process itself, such as quantization-based methods~\cite{jegouProductQuantizationNearest2011, DBLP:journals/pami/GeHK014,gao2019beyond, guoAcceleratingLargeScaleInference2020,DBLP:journals/pami/GeHK014}.

Our method belongs to the first category and is orthogonal to the second category. Our proposed index differs from previous ANNS indexes in 
two aspects: (1) Our proposed index is designed for embeddings based spatial keyword queries. In contrast, existing ANNS indexes are designed for embedding retrieval and do not work well when being adapted to embedding based spatial keyword queries (to be shown in Section~\ref{sec:exp-effectiveness}). 
(2) Existing ANNS techniques are based on heuristic methods. If we want to incorporate spatial factor into the existing ANNS techniques, it is unavoidable to balance the weight between spatial and embedding relevance, which is however challenging. To avoid this, our proposed index employs neural networks to learn 
to cluster relevant geo-textual objects together while separating the irrelevant ones, and routes embedding based spatial keyword queries to relevant clusters through the neural networks to accelerate the top-k search process. 
Notably, our proposed index utilizes a novel learning-to-cluster technique to group embedding based geo-textual objects, while the IVF index employs the conventional $k$-means algorithm to cluster embeddings.

\section{Problem Statement and Motivations}
\label{sec:problem-statement-and-motivations}
\subsection{Preliminary}
\label{sec:preliminary}
We consider a geo-textual object dataset $D$, where each geo-textual object $o\in D$ has a location description $o.loc$ and a textual description $o.doc$. The location description $o.loc$ is a two-dimensional GPS coordinate composed of latitude and longitude. The textual description $o.doc$ is a document that describes the object. The Top-k KNN Spatial-Keyword Query (\tkskq)~\cite{congEfficientRetrievalTopk2009} is defined as follows.

\noindent\textbf{Top-k KNN Spatial-Keyword Query (TkQ):} Given a query $q=\langle loc, doc, k\rangle$, where $q.doc$ denotes the query keywords, $q.loc$ is the query location, and $q.k$ is the number of returned objects, we aim to retrieve $k$ objects with the highest relevance scores $ST(q,o)$:
\begin{equation}
\begin{aligned}
    ST(q,o) = (1-\alpha) \times SRel(q.loc, o.loc) + \\
    \alpha \times TRel(q.doc, o.doc).
\end{aligned}
\label{eq:TkSKQ_rank}
\end{equation}
A higher score $ST(q,o)$ indicates higher relevance between the given query $q$ and object $o$. In this context, $SRel(q.loc, o.loc)$ denotes the spatial relevance between $q.loc$ and $o.loc$ and is often calculated by $1-SDist(q.loc, o.loc)$ in previous studies~\cite{congEfficientRetrievalTopk2009,DBLP:conf/icde/FelipeHR08,DBLP:journals/tkde/LiLZLLW11, DBLP:conf/ssd/RochaGJN11,liu2023effectiveness}, where $SDist(q.loc, o.loc)$ represents the spatial closeness and is usually computed by the normalized Euclidean distance: $SDist(q.loc, o.loc)=\frac{dist(q.loc, o.loc)}{dist_{max}}$. Here, $dist(q.loc, o.loc)$ denotes the Euclidean distance between $q.loc$ and $o.loc$, and $dist_{max}$ is the maximum distance between any two objects in the object dataset. $TRel(q.doc, o.doc)$ denotes the text relevance between $p.doc$ and $q.doc$ and is computed by traditional information retrieval models like BM25~\cite{DBLP:journals/ftir/RobertsonZ09} in previous studies~\cite{congEfficientRetrievalTopk2009,DBLP:conf/icde/FelipeHR08,DBLP:journals/tkde/LiLZLLW11, DBLP:conf/ssd/RochaGJN11}, and then normalize to a scale similar to spatial relevance. $\alpha \in [0,1]$ is a weight parameter to balance the spatial and text relevance.

\subsection{Data Analysis and Motivations}
\label{sec:motivations}

\begin{figure}[!t]
\centering
\begin{subfigure}[b]{0.45\columnwidth}
    \centering
    \includegraphics[width=\textwidth]{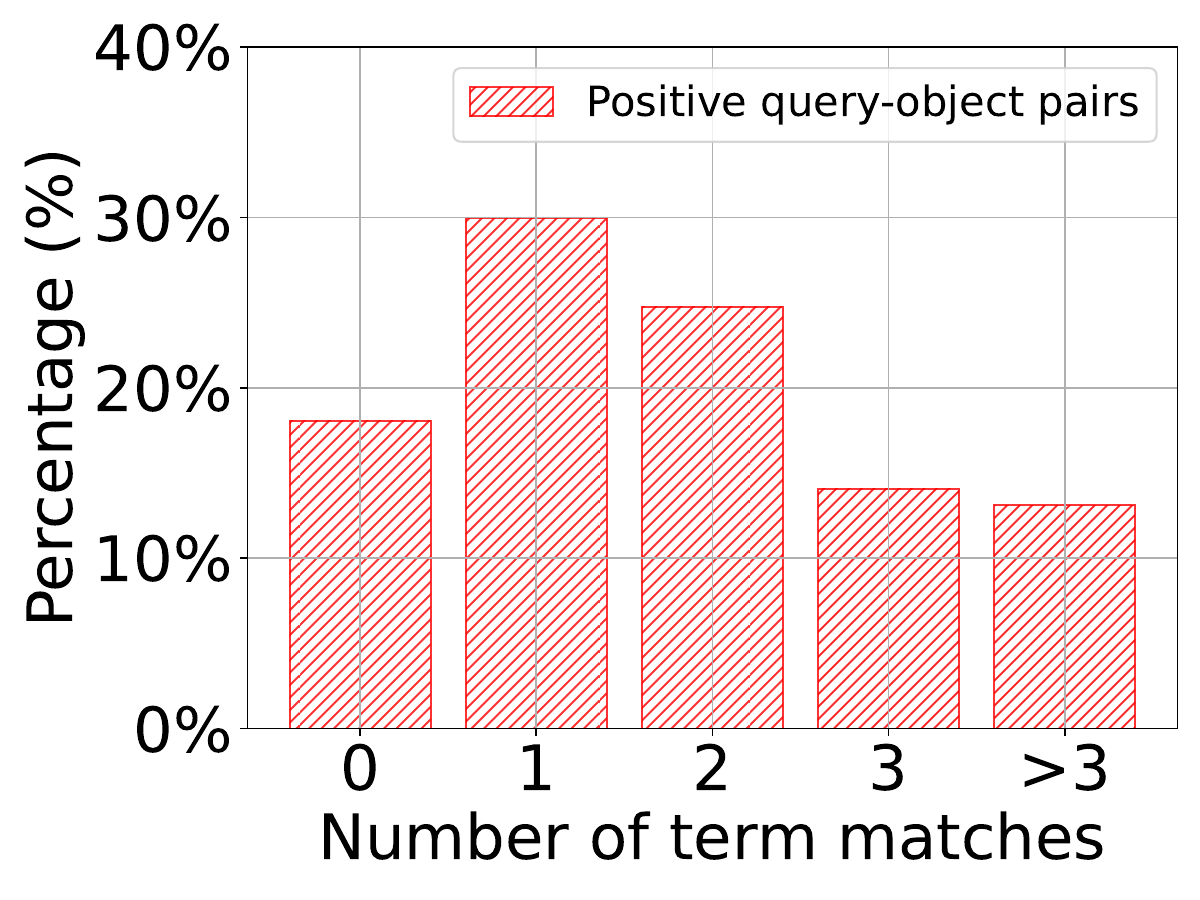}
    \caption{Word Mismatch Analysis}
    \label{fig:term-match}
\end{subfigure}
\hfill
\begin{subfigure}[b]{0.45\columnwidth}
    \centering
    \includegraphics[width=\textwidth]{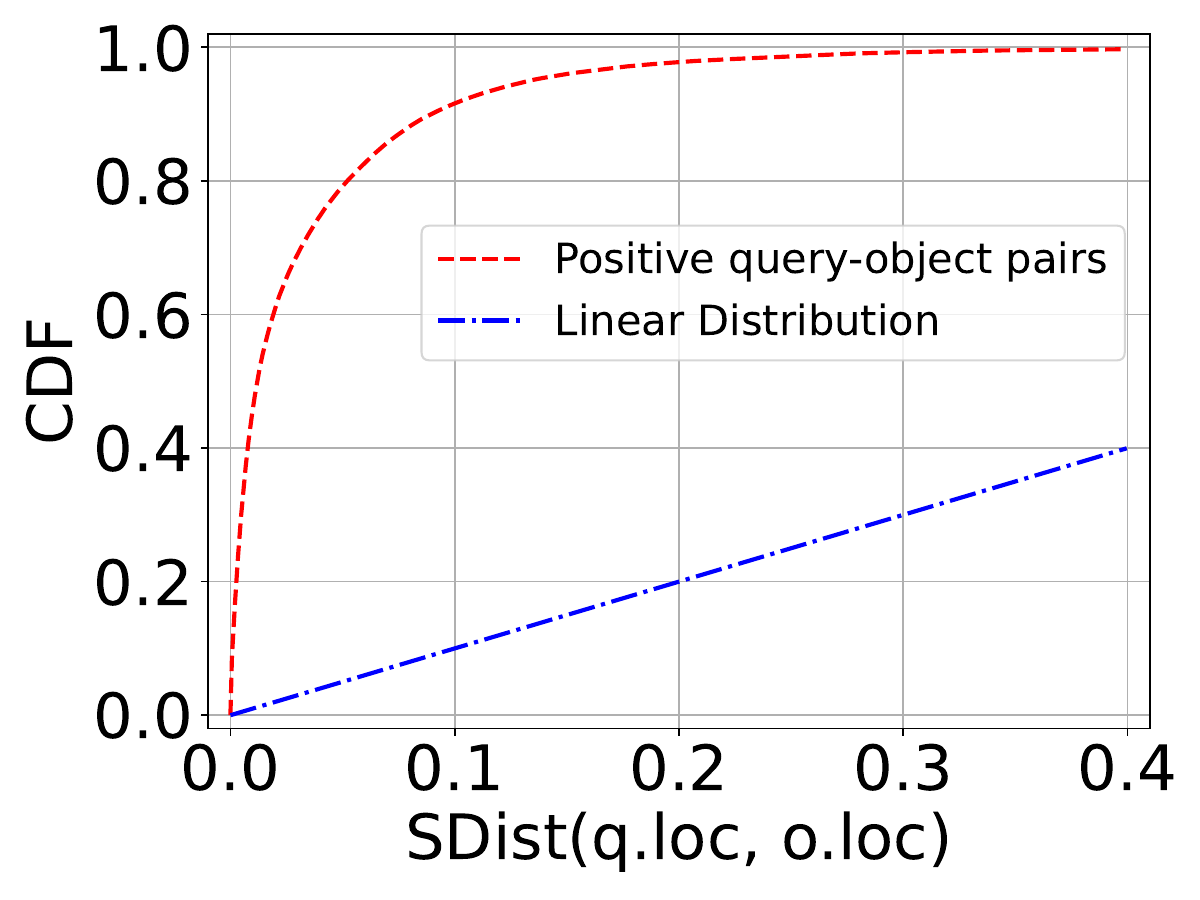}
    \caption{Spatial Relevance Analysis}
    \label{fig:cdf_distribution}
\end{subfigure}
\caption{Figure~\ref{fig:term-match} shows the percentage distribution of ground-truth positive query-object pairs based on the number of matching terms on the Beijing dataset. Figure~\ref{fig:cdf_distribution} compares the CDF of spatial distance for ground-truth positive query-object pairs and the linear distribution on the Beijing dataset.}
\label{fig:spatial_relevance_score}
\end{figure}

\noindent\textbf{Word Mismatch.} As discussed in Section~\ref{sec:intro}, \tkskq~\cite{congEfficientRetrievalTopk2009,DBLP:conf/icde/FelipeHR08,DBLP:conf/ssd/RochaGJN11} typically uses traditional information retrieval methods such as BM25 and TF-IDF~\cite{DBLP:journals/ftir/RobertsonZ09,DBLP:journals/tois/WuLWK08} to compute textual relevance. These models rely on exact word matching to compute textual relevance and thus suffer from the word mismatch issue, which reduces their effectiveness. We further illustrate this issue in the real-world dataset that contains query-object ground-truth pairs. The details on the data are given in Section~\ref{sec:exp}. As shown in Figure~\ref{fig:term-match}, in the Beijing dataset, nearly 20\% of the ground-truth query-object positive (relevant) pairs have no overlap of words. This indicates that objects relevant to the query may receive a low or even zero textual relevance from these models. For instance, given a search `nearby drugstore', if the search engine employs the BM25 to determine textual relevance, a drugstore labeled as `pharmacy' would get a textual relevance score of zero, although the two terms convey the same meaning.

\noindent\textbf{Efficiency.} To address the word mismatch issue, several recent studies~\cite{liu2023effectiveness,DBLP:conf/aaai/ZhaoPWCYZMCYQ19,DBLP:conf/sigir/YuanLLZYZX20} have utilized deep learning techniques to evaluate textual relevance, thereby enhancing ranking effectiveness. However, these methods rely on word embeddings and complex neural networks to compute textual relevance, resulting in high querying latency. For instance, DrW~\cite{liu2023effectiveness} identifies the top-k relevant words from each object's description for each query word based on their word embeddings, and then aggregates the scores of these word pairs using an attention mechanism to compute the textual relevance. On the Geo-Glue dataset, which comprises 2.8 million objects, DrW takes more than 7 seconds to answer a query on average in our experiment, which aligns with the results reported in~\cite{liu2023effectiveness}. This makes them unsuitable as a retriever for practical geo-textual object retrieval applications, although they can be used as re-rankers for a small number of objects returned by a retriever.

\noindent\textbf{Motivations of \Framework.} As discussed in the last paragraph, incorporating deep textual relevance into spatial keyword queries has presented significant efficiency challenges. However, as discussed in Section~\ref{sec:related}, the first category of deep text relevance model, namely word interaction based deep relevance models, such as DrW~\cite{liu2023effectiveness}, which rely on word interaction modules and word embeddings to compute textual relevance, are not efficient for online computation and retrieval by a specifically designed index. Therefore, \revision{we consider embeddings based spatial keyword queries, which capture the semantic meaning of query keywords and object descriptions into two separate embeddings. Then the challenge lies in how to develop an index for embedding based spatial keyword queries as discussed in Introduction.}


\noindent\textbf{Spatial Relevance.} 
\revision{Previous studies, such as TkQ~\cite{congEfficientRetrievalTopk2009} and DrW~\cite{liu2023effectiveness}, typically compute spatial relevance by $1-SDist(q.loc,o.loc)$.}
The implicit assumption behind this linear function is that the user's geographical preference for geo-textual objects is a linear function of distance. However, this assumption was not examined previously and we found it does not hold on our real-life datasets. In Figure~\ref{fig:cdf_distribution}, we illustrate this issue by comparing the cumulative distribution function (CDF) of ground-truth positive (relevant) query-object pairs with the `Linear Distribution', a linearly ascending hypothetical scenario that positive query-object pairs are uniformly distributed across the range $[0,1]$. This figure shows a sharp increase in the CDF of ground-truth positive pairs for $SDist(q.loc,o.loc)$ below 0.1, a pattern that 
greatly differs from the hypothetical scenario. This motivates us to design a new spatial relevance module to better fit the real-world scenario. An intuitive solution would be to employ an exponential function. However, our experiments indicate that this method is even less effective compared to the linear function (as to be shown in Section~\ref{sec:exp-ablation}).

\noindent\textbf{Problem Statement.} \revision{ We aim to develop an ANNS index for deep learning based relevance models designed for embedding based spatial keyword queries, where query keywords and object descriptions are represented as two separate embeddings. 
}


\section{Proposed Retriever (\Framework)}
\label{sec:LIST}
\begin{figure}[t]
\centering
\includegraphics[width=0.45\textwidth]{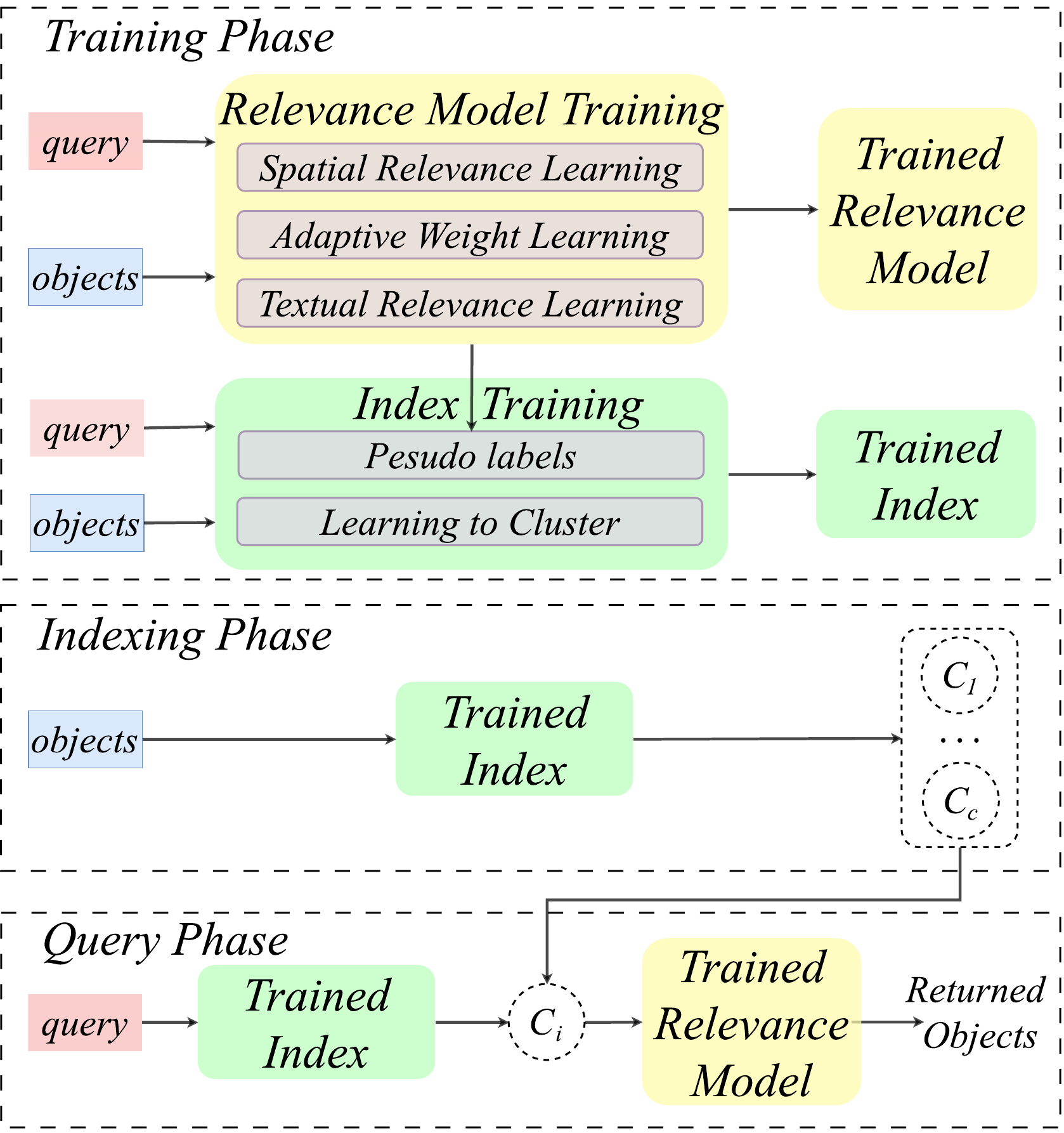}
\caption{The three phases of our retriever \Framework: the training, indexing, and query phase. 
The relevance model is shown in yellow and the index is shown in green.}
\label{fig:framework}
\end{figure}

\subsection{Overview}
\label{sec:overview}
\revision{\Framework is featured with a relevance model and a 
machine learning based ANNS index. Notably, our proposed index is applicable to any other relevance model developed for embedding based spatial keyword queries.} 

The workflow of \Framework is summarized in Figure~\ref{fig:framework}, which has three phases: training, indexing, and query phase. During the training phase, we first train the relevance model, and then train our index. During the indexing phase, each object is partitioned into one cluster. During the query phase, the trained index routes each query to either a single cluster or a subset of clusters that have the highest probabilities. Within these clusters, \Framework returns $k$ objects with the highest scores ranked by the trained relevance model as the query result.

\subsection{The  Relevance Model}
\label{sec:model}
\noindent\textbf{Textual Relevance Learning.} Inspired by the dual-encoder model's success in document retrieval~\cite{DBLP:conf/emnlp/KarpukhinOMLWEC20}, we employ a dual-encoder module to encode query keywords and object description into two separate embeddings, 
and calculate textual relevance by the inner product between the two embeddings. 
Compared with previous relevance models~\cite{liu2023effectiveness} that rely on word embeddings and complex word interaction functions discussed in Section~\ref{sec:related}, 
the dual encoder module is both efficient and effective.



The dual-encoder module comprises an object encoder $E_o$ and a query encoder $E_q$, \revision{each of which is a BERT model~\cite{devlin2018bert}.} The encoder takes the textual content of the query or object as input, captures interactions between words by the transformer-based model, and finally utilizes the representation of the [CLS] token as the global semantic representation, which is a $d$-dimensional embedding. This process is formulated as:
\begin{equation}
\begin{aligned}
    o.emb = E_{o}(o.doc; \theta_o), \: o.emb \in \mathbb{R}^{d}, \\
    q.emb = E_{q}(q.doc; \theta_q), \: q.emb \in \mathbb{R}^{d}.     
\end{aligned}
\label{eq:DE}
\end{equation}
where $E_{o}(·; \theta_o)$ denotes the object encoder parameterized with $\theta_o$ and $E_{q}(·; \theta_q)$ denotes the query encoder parameterized with $\theta_q$. Then the text relevance score is calculated by the inner product between the $q.emb$ and $o.emb$,
\begin{equation}
TRel(q.doc,o.doc) = q.emb \cdot o.emb.
\label{eq:textual-relevance}
\end{equation}

{\color{red} }




\noindent\textbf{Spatial Relevance Learning.} As discussed in Section~\ref{sec:motivations}, it is essential to develop a more effective spatial relevance module. Therefore, we propose a new learning-based spatial relevance module to learn to estimate spatial relevance.




As discovered in Section~\ref{sec:motivations}, the user's geographic preferences for geo-textual objects do not follow a linear pattern. \revision{Here, we consider two features of the 
preference: 
(1) it increases as the distance decreases, and (2) it exhibits a stepwise decline as the distance increases. The first feature is straightforward because users tend to visit nearby objects. Then a straightforward solution is to design a monotonically continuous function to learn from query-object positive pairs and predict users' spatial preferences. However, this approach leads to overfitting. For example, if most positive pairs within training dataset are distributed within 1km, the continuous function will learn 1km as a boundary, causing high spatial relevance within 1km but a sharp drop beyond it. Actually, 1.1km is not significantly different from 1km for users. This overfitting results in poor performance due to lack of generalization (as to be shown in Section~\ref{sec:exp-ablation}). To alleviate this issue, we propose the second feature. To explain this stepwise pattern, let us consider a scenario in which a customer wishes to purchase coffee from Starbucks. If the nearest Starbucks is very close, s/he would go and buy coffee. If the nearest Starbucks is a little far, s/he may hesitate and her/his intention of visiting the Starbucks would decrease. If the nearest Starbucks is very far away, s/he would give up this idea. Therefore, this pattern aligns with real-world characteristics and alleviates the overfitting issue as the boundaries are manually based on our knowledge. In summary, the new spatial relevance module is designed to be a learnable monotonically step function.}

Our proposed spatial relevance module takes $S_{in}=1-SDist(q.loc,o.loc)$ as input, consisting of a threshold array $\mathcal{T} \in \mathbb{R}^{t \times 1}$ and a learnable weight array $w_s \in \mathbb{R}^{1 \times t}$. Here, $\mathcal{T}$ stores the threshold values that determine the transition points of the step function, i.e., the value exceeded by $S_{in}$ will trigger an increase of spatial relevance, which is used to ensure that the learned function is a step function. Specifically, $\mathcal{T}$ is structured as $\mathcal{T}[i] = \frac{i}{t}$, where $i \in [0,t]$ and $t$ is a hyperparameter to control the increment of the threshold value. For example, when $t=100$, $\mathcal{T}$ is $[0.0,0.01,\cdots, 0.99,1.0]$. The learnable weight array $w_s$ determines the extent of the increase when the input $S_{in}$ reaches these threshold values, which are learned from the training data and used to estimate the spatial relevance. When the input $S_{in}$ exceeds the value of $\mathcal{T}[i]$, then the spatial relevance increases by $act(w_s[i])$, where $act$ is an activation function to ensure $act(w_s[i])$ remain non-negative. This process ensures that the output $SRel$ exhibits a step increase as the input $S_{in}$ increases. The learned spatial relevance is computed as below: 
\begin{equation}
    SRel(q.loc,o.loc) = act(w_{s})  \cdot \mathbb{I}(S_{in} \geq \mathcal{T}[i]),
\label{eq:DSR}
\end{equation}
where $SRel(q.loc,o.loc)$ is the learned spatial relevance. $\mathbb{I} \in \{0,1\}^{t \times 1}$ is an indicator array. $\mathbb{I}[i]=1$ if $S_{in} \geq$ $\mathcal{T}[i]$; otherwise 0. The sum of the step increment is conducted by an inner product between the indicator array $\mathbb{I} \in \{0,1\}^{t \times 1}$ and the learnable weight after activation $act(w_{s}) \in \mathbb{R}^{1 \times t}$.



During the query phase, we extract the weights in $w_{s}$ from the module and store them as an array $\hat{w_{s}}$ for faster inference. $\hat{w_{s}}$ is constructed as $\hat{w_{s}}[i] = \sum\limits_0^{i}act(w_{s}[i])$. When computing the spatial relevance, we get the input $S_{in}$. Since the threshold value grows uniformly by $\frac{1}{t}$, we can determine the number of threshold values exceeded by the input as $\lfloor S_{in}*t \rfloor$, where $\lfloor . \rfloor$ indicates a floor function and is utilized to truncate a real number to an integer. This also corresponds to the sum of the values of weights, which is the spatial relevance score. This process is formulated as:
\begin{equation}
SRel(q.loc,o.loc) = \hat{w_{s}}[\lfloor S_{in}*t\rfloor].
\label{eq:DSR_inference}
\end{equation}
Hence, during the query phase, the time complexity of computing spatial relevance is O(1), which is efficient.

\noindent\textbf{Adaptive Weight Learning.} The recent study~\cite{liu2023effectiveness} has shown the importance of weight learning in improving ranking effectiveness. To enhance ranking effectiveness, we propose an adaptive weight learning module, which aims to assign adaptive weights to textual and spatial relevance based on the query keywords. \revision{For instance, in the case of detailed query keywords, such as `gas station nearby', giving lower weight to textual relevance better fits the real-world scenario. Since there may be many gas stations nearby, users tend to prefer the closest one. Conversely, in another scenario, such as searching with keywords `Lincoln Memorial in Washington, D.C.', due to the uniqueness of the keyword, giving greater weight to textual relevance can deliver better results.}

Hence, we employ a simple yet effective manner that directly utilizes an MLP layer to determine the weights based on the embedding of query keywords $q.emb$, which is formulated as follows:
\begin{equation}
    w_{st} = MLP(q.emb), \: w_{st}\in \mathbb{R}^{1\times 2}.
\label{eq:model_weight}
\end{equation}

Similar to the Equation~\ref{eq:TkSKQ_rank}, the final relevance score between the query and the object is calculated as:
\begin{equation}
    ST(q,o) = w_{st}\cdot [TRel(q.doc, o.doc), SRel(q.loc,o.loc)]^{T}. 
\label{eq:model_rank}
\end{equation}

\noindent\textbf{Training Strategy.} To train our relevance model, we employ the 
contrastive learning strategy. Given a query $q_i$, the positive (relevant) geo-textual object $o_i^{+}$ is obtained by real-world 
ground-truth data (detailed in Section~\ref{sec:exp-setup}). As for the negative (irrelevant) geo-textual objects, previous studies of passage retrieval~\cite{DBLP:conf/naacl/QuDLLRZDWW21,renRocketQAv2JointTraining2021} have shown that hard negative objects can improve retrieval performance. Inspired by them, we introduce this mechanism for spatial keyword query, choosing a subset of hard negative objects for training. Specifically, for each training query, we first filter out the positive objects and then use TkQ~\cite{congEfficientRetrievalTopk2009} to retrieve a set of top-ranked objects, which are considered the hard negative set for that query. In each training epoch, we randomly pick $b$ hard negative samples from this set. We optimize the loss function as the negative log-likelihood of the positive object: 

\begin{equation}
\begin{split}
    L_{\text{model}} & (q_i, o_i^{+}, o_{i,1}^{-}, o_{i,2}^{-}, \cdots, o_{i,b}^{-}) \\
    & = -log \frac{e^{ST(q_i, o_i^{+})}}{e^{ST(q_i, o_i^{+})} + \sum_{j=1}^{b}e^{ST(q_i, o_{i,j}^{-})}}.
    \label{eq:loss}
\end{split}
\end{equation}
In addition, we also utilize the in-batch negatives strategy~\cite{lin2021inbatch} to further enhance training efficiency.

\noindent\revision{\textbf{Complexity Analysis.} Now we analyze the time 
complexity of the proposed relevance model. Assuming the dimensionality of the embedding is $d$, the total number of objects is $n$, and the embeddings of objects are generated in advance. The time complexity of a brute-force search of our relevance model over the entire dataset is $O(n(d + 2) + d)$. Here, $O(nd)$ represents the time cost of Equation~\ref{eq:textual-relevance}, $O(2n)$ corresponds to the time cost of Equation~\ref{eq:DSR_inference}, and $O(d)$ accounts for the time cost of Equation~\ref{eq:model_weight}.} 

\revision{When compared with other relevance models like DrW~\cite{liu2023effectiveness}, which has a time complexity of $O(nl^q(l^od+k')+l^q(2d^2+d))$, our relevance model exhibits superior efficiency in terms of time complexity. Here, $l^q$ represents the number of words in the query keywords, $l^o$ represents the number of words in the object description, and $k'$ denotes the number of top-k' relevant terms to find in the object description for each query keyword. These hyperparameters, $l^q$, $l^o$, and $k'$, are typically in the dozens. Consequently, DrW is significantly slower than our relevance model.}

\begin{figure}[t]
  \centering
  \includegraphics[width=0.45\textwidth]{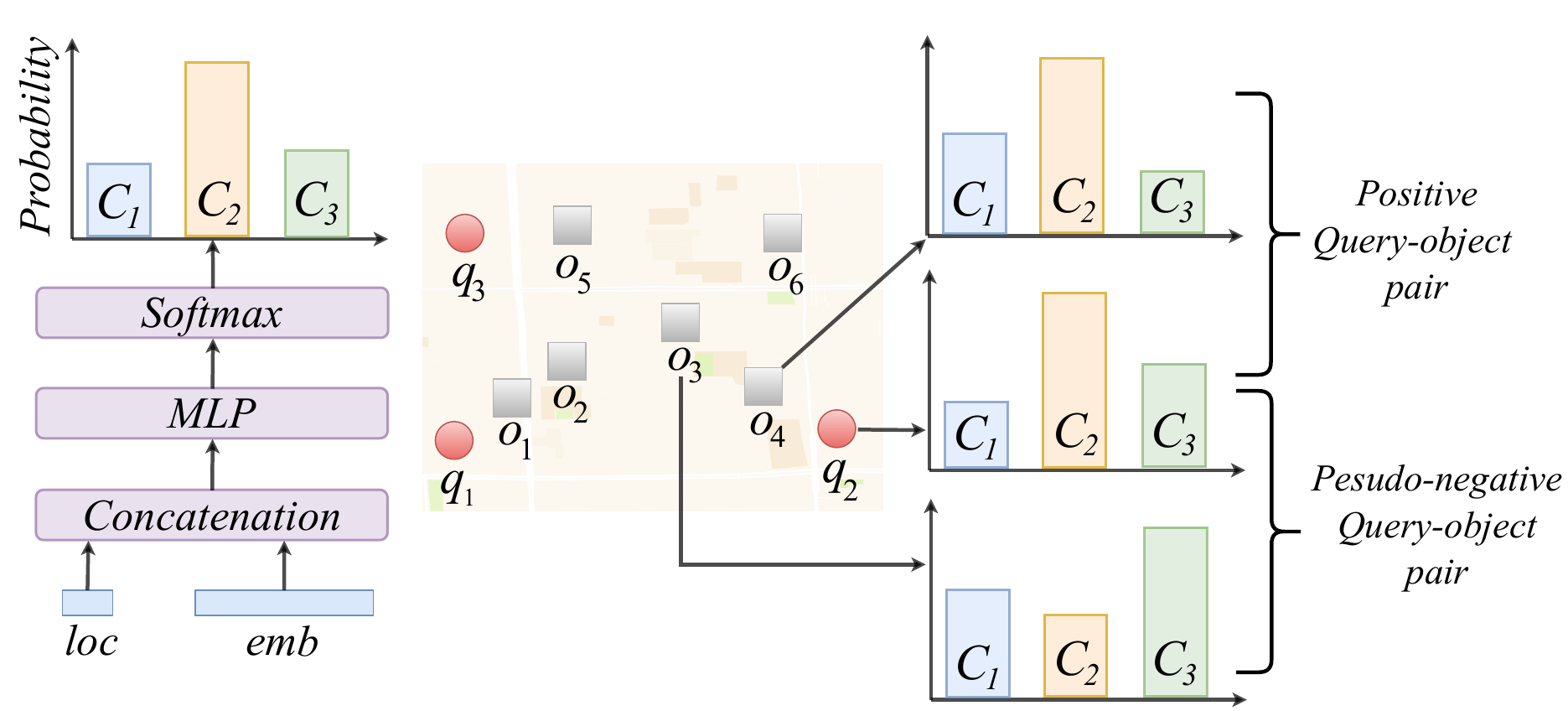} 
  \caption{The illustration of the index.}
  \label{fig:index}
\end{figure}

\subsection{The Proposed Index}
\label{sec:index}
In this section, our objective is to develop a new ANNS index that, for each embedding based spatial keyword query, can return a subset of objects that receive high scores from the proposed relevance model for the input query, thereby accelerating the top-k search process. 


To build such an index, 
our proposed index adopts a machine learning based method. Without manually setting the weight between spatial and textual relevance, it learns to cluster relevant queries and objects together while separating the irrelevant ones. Specifically, as illustrated in Figure~\ref{fig:index}, it takes the geo-location and textual embeddings of objects or queries as input and employs a Multi-Layer Perceptron (MLP) to partition objects and route queries into $c$ clusters, and then searches for the top-k relevant objects within the routed cluster. To make the index both effective and efficient, we develop a novel pseudo-negative query-object pair generation method. These pseudo labels combined with the positive (relevant) query-object pairs ensure that relevant queries and objects are partitioned and routed into the same cluster while separating the irrelevant ones, thereby reducing the search space while maintaining effectiveness.

\noindent\textbf{Feature Construction.} 
We utilize geo-location and textual embedding to construct a consistent input representation for both objects and queries, which is illustrated in Figure~\ref{fig:index}. \revision{The textual embedding $emb \in \mathbb{R}^d$ is a $d$-dimensional embedding converted from textual content $o.doc$ or $q.doc$ by the trained dual encoder module, which is typically hunderds, e.g., 768. It is L2 normalized before being input into the neural network.} The geo-location $loc=\langle lat,lon \rangle$ are transformed into the following features $\langle \hat{lat}, \hat{lon} \rangle$ as below:  
\begin{equation}
\hat{lat} = \frac{lat-lat_{min}}{lat_{max}-lat_{min}}, \hat{lon} = \frac{lon-lon_{min}}{lon_{max}-lon_{min}},
\label{eq:normalize}
\end{equation}
where $lat_{min}$ ($lon_{min}$) represents the lowest latitude (longitude) in dataset $D$, $lat_{max}$ ($lon_{min}$) denotes the highest latitude (longitude) in dataset $D$. Then the input representation is formulated as 
\begin{equation}
x = [emb, \hat{lat}, \hat{lon}].
\label{eq:representation}
\end{equation}
Here $x$ is the input feature vector for the cluster classifier, i.e., the MLP. \revision{Since the two feature vectors are normalized independently, they will have a similar impact on the neural network.}

\noindent\textbf{Cluster Classifier.} Our cluster classifier is a lightweight Multi-Perception Layer (MLP). Given $x$ as the input, it produces the $c$-cluster probability distribution, defined as follows:
\begin{equation}
    Prob = Softmax(MLP(x)), \: Prob \in \left[0, 1\right]^{c \times 1},
    \label{eq:L2PO}
\end{equation}
where $c$ is a hyperparameter that indicates the desired number of clusters we aim to obtain, and $Prob$ represents the predicted $c$-cluster probability distribution. The $c$-cluster probability distribution of object $o$ is denoted as $Prob_o$ and that of query $q$ is represented as $Prob_q$. 

Note that the cluster classifier, i.e., the MLP, is shared between queries and objects. This setting allows our index to learn the distribution of both queries and objects, subsequently grouping identified relevant queries and objects into the same cluster (as detailed later).




\noindent\textbf{Pseudo-Label Generation.} To train the cluster classifier, both positive and negative pairwise labels are required. We utilize the ground-truth query-object relevant label (e.g., click-through or human annotation, detailed in Section~\ref{sec:exp-setup})  as the positive pairwise label for training. However, we lack high-quality negative pairwise labels. Randomly selecting negative objects will lead to overfitting and all objects are grouped together (as detailed later). Therefore, we propose a novel pseudo pairwise negative label generation method. 

The relationship between query $q$ and object $o$ is denoted as $s(q,o)$. If $q$ and $o$ are relevant 
, $s(q,o)=1$; otherwise, $s(q,o)=0$, which are then used as labels for training. We leverage ground-truth positive labels as pairwise positive labels in the training, as shown below: 
\begin{equation}
    \text{pos}_{q} = \{o; o\in D, s(q,o)=1\},
    \label{eq:positive_pool}
\end{equation} 
where the positive object set of query $q$ is denoted as $\text{pos}_{q}$.

The negative object set $\text{neg}_{q}$ of query $q$ is generated by the relevance model. Given a query $q$, we employ the relevance model to calculate the relevance score for all objects. Then we adaptively select $\text{neg}_{q}$ according to two hyperparameters $neg_{start}$ and $neg_{end}$, as shown below:
\begin{equation}
    \text{neg}_{q} = \text{argsort}_{o \in D} ST(q,o)[neg_{start} : neg_{end}], \; s(q,o)=0,
    \label{eq:negative_pool}
\end{equation} 
where $ST(q,o)$ is the relevance score between query $q$ and object $o$ produced by the relevance model. Note that positive query-object pairs are excluded as indicated by the filter condition $s(q,o)=0$. 

This adaptive pseudo-negative generation method draws inspiration from hard negative sample strategy~\cite{DBLP:conf/naacl/QuDLLRZDWW21,renRocketQAv2JointTraining2021}. What sets our approach apart from existing studies~\cite{DBLP:conf/naacl/QuDLLRZDWW21,renRocketQAv2JointTraining2021} is that we employ an adaptive manner to select hard negative samples, thereby controlling the difficulty level of the generated negative samples. This adjustment strikes the trade-off between the effectiveness and efficiency. 

Decreasing $neg_{start}$ leads to a set of harder negative objects being chosen to train the model. Consequently, the classifier is more effective in distinguishing between positive objects and hard negative objects, which facilitates the clustering of relevant queries and objects, while effectively segregating the irrelevant ones. Empirically, under this setting, only the very relevant queries and objects are grouped into the same cluster. The reduced number of objects within a query's cluster leads to higher efficiency. In contrast, when using a large $neg_{start}$, the index can not learn useful information, leading to a scenario where all objects tend to be clustered together.

On the other hand, when our index is trained on a set of harder negative objects, it might also exclude some positive ones and reduce its effectiveness. Thus, the choice of $neg_{start}$ strikes a trade-off between effectiveness and efficiency. This alleviates the lack of negative label issue and the skewed cluster distribution issue of existing techniques~\cite{Hsu16_KCL,Hsu18_L2C}.

\noindent\textbf{Training Strategy.} Based on the hard negative objects and positive objects provided above, we employ the MCL loss function~\cite{Hsu19_MCL} to train the MLP, as described below: 
\begin{equation}
\begin{split}
L_{\text{Index}} & (q_i, o_i^{+}, o_{i,1}^{-}, o_{i,2}^{-}, \cdots, o_{i,m}^{-})\\
    & = log(\hat{s}(q_i, o_i^{+})) + \sum\limits_{j=1}^{m} log(1-\hat{s}(q_i,o_{i,j}^{-})),
\end{split}
\label{eq:MCL}
\end{equation}
where $\hat{s}(q_i,o_j) = Prob^{T}_{q_i}\cdot Prob_{o_j}$, $o_i^{+} \in \text{pos}_q$ and $o_{i,j}^{-} \in \text{neg}_q$. Typically, we randomly select one positive object $o_i^{+}$ from the positive object set $\text{pos}_q$ and $m$ negative objects from the negative object set $\text{neg}_q$ in each training epoch. 


As described earlier, the MLP is shared between queries and objects. Through this training process, for positive query-object pairs, their $Prob_o$ and $Prob_q$  will have a similar distribution. Thereby, relevant pairs of query and object are more likely to be grouped into the same cluster while the irrelevant pairs are grouped into distinct clusters. For example, as illustrated in Figure~\ref{fig:index}, the object $o_4$ and query $q_2$ form a positive query-object pair, and they are expected to be grouped into the same cluster $C_2$.


\noindent\textbf{Learning to Partition and Route.} During the indexing phase, each object $o$ is partitioned to the cluster with the highest probability according to $Prob_o$. 
Once the partitioning is completed, the objects assigned to a cluster are stored in a corresponding list, which acts as an inverted file for these objects. Each object in this list is represented by a $d$-dimensional vector $o.emb$, and a geo-location $o.loc$, which will be utilized to calculate the relevance score for incoming queries. During the query phase, a given query $q$ is directed to the cluster that has the highest probability. Subsequently, the relevance scores between $q$ and all objects within the cluster are calculated and the top-k objects are selected as result objects.

An alternative way is routing queries (objects) to $cr$ clusters with the highest probabilities based on $Prob_q$ ($Prob_o$). Although this might boost query effectiveness by considering more objects in different clusters, it sacrifices efficiency as more objects need to be calculated for a query, leading to an accuracy-efficiency trade-off. 

\noindent\textbf{Insertion and Deletion Policy.} When a new object comes, we convert it into an embedding using the trained relevance model and then assign it to specific clusters using the trained index. 
When an object is deleted, we simply remove it from the corresponding cluster. \revision{The time cost of the insertion operation is equivalent to the inference time of the neural networks, while the time cost of the deletion operation involves scanning the cluster lists and then deleting the object's id. The time costs of both operations are negligible.}

\revision{Note that most existing ANNS indexes are static~\cite{luoSurveyDeepHashing2023,DBLP:journals/pami/WangZSSS18,DBLP:journals/pvldb/WangXY021}. When inserted data significantly differs in distribution from the existing data or when insertions occur frequently, periodically rebuilding the index is necessary to maintain high accuracy. This is because most ANNS indexes are based on the principle of organizing similar objects together, thereby routing queries to similar blocks to reduce search space. The functions determining which cluster each object belongs to are typically static, such as the distance between objects and the centroids of each cluster in an IVF index. When the distribution of embeddings changes significantly, these static functions can no longer accurately capture the new distribution, leading to a drop in performance. In our approach, we address this issue by only retraining the index, but not the relevance model.} 





\noindent\textbf{Cluster Evaluation.} After the training phase, we employ the trained index to produce clusters $C=\{C_1,C_2,\\ \cdots,C_{c}\}$, where $C_i$ represents cluster $i$. We use validation queries to evaluate the quality of the clusters. The validation queries are fed into the trained index and routed to a cluster. $C_{i}^{q}$ denotes a list containing the validation queries routed to $C_i$, $C_{i}^{o}$ denotes a list containing objects partitioned to $C_i$, and $||$ indicates the size of a list. We proceed to introduce two metrics to evaluate the quality of the clusters.

The first metric evaluates the precision of $C_i$, denoted as $P(C_i)$, representing the degree to which queries are aligned with their corresponding positive (relevant) objects in the same cluster, which is defined below: 
\begin{equation}
P(C_{i}) = \frac{1}{|C_{i}^{q}|} \sum\limits_{q_j \in C_{i}^{q}}\frac{|\text{pos}_{q_j}\cap C_{i}^{o}|}{|\text{pos}_{q_j}|}.
\label{eq:precision_cluster}
\end{equation}
Building upon this, we compute the average precision across all clusters, denoted as $P(C)$, which is defined as below: 
\begin{equation}
P(C) = \frac{1}{\sum\limits_{C_{i} \in C}|C_{i}^{q}|} \sum_{C_i \in C} P(C_i) * |C_i^{q}|.
\label{eq:precision}
\end{equation}
Intuitively, a higher $P(C)$ indicates that the index is more effective.

In addition, we also take into account efficiency concerns. For this purpose, we introduce another metric, the Imbalance Factor (IF)~\cite{johnson2019billion}, which measures the degree of balance across all clusters, denoted as $\text{IF}(C)$. The $IF(C)$ is formulated as: 
\begin{equation}
\text{IF}(C) = \frac{\sum|C_{i}|^2}{(\sum|C_{i}|)^2},
\label{eq:IF}
\end{equation}
where $\text{IF}(C)$ is minimized when $|C_1|=|C_2|=\cdots=|C_{c}|$ according to the Cauchy-Schwarz Inequality. A higher imbalance factor indicates a more uneven distribution of clusters. When most objects are concentrated in a few large clusters, the imbalance factor increases significantly, which is undesirable for our task. Overall, our goal is to achieve high-quality clustering results characterized by higher $P(C)$ and lower $\text{IF}(C)$. In our index, with proper hyperparameters, our training process of the cluster classifier is able to obtain high-quality clusters. In the experiments, We study the quality of the clusters produced by our proposed index using these evaluation metrics 
(shown in Section~\ref{sec:exp-ablation}).








\subsection{Procedures and Analyses of \Framework}
\label{sec:workflow}
\begin{algorithm}[t]
    \caption{Procedures of LIST.}
	\label{alg:workflow}
	\KwIn{A geo-textual object dataset $D$, a relevance model $R$, a index $I$, training \tkskqs set $Q_{\text{train}}$, and incoming \tkskqs set $Q$}
	\KwOut{The response $Res_{q}$ for each query $q \in Q$}
    \vspace*{1ex}
	// \textbf{Training Phase: } $Train(Q_{\text{train}}, D, R, I)$\\
    Train $R$ by $Q_{\text{train}}$, $D$ based on Eqaution~\ref{eq:loss}\;
    Employ $R$ to generate pseudo-labels based on Equation~\ref{eq:negative_pool}\;
    Employ pseudo-labels to train $I$ based on Equation~\ref{eq:MCL}\;
    \vspace*{1ex}
	// \textbf{Indexing Phase: } $Indexing(D)$\\
    \For{$o \in D$}{
        Transform $o.doc$ to $o.emb$ based on Equation~\ref{eq:DE}\;
        Transform $o.emb$, $o.loc$ to $x_o$ based on Equation~\ref{eq:representation}\;     
        Generate $Prob_{o}$ by $MLP$ based on Equation~\ref{eq:L2PO}\;
        Parition $o$ to cluster $C_i$ based on $Prob_{o}$\;
    }
    \vspace*{1ex}
    // \textbf{Query Phase: } $Search(q, C, R, I)$\\
    \For{$q \in Q$}{
        Transform $q.doc$ to $q.emb$ based on Equation~\ref{eq:DE}\;
        Transform $q.emb$, $q.loc$ to $x_q$ based on Equation~\ref{eq:representation}\;     
        Generate $Prob_{q}$ by $I$ based on Equation~\ref{eq:L2PO}\;        
        Route $q$ to cluster $C_i$ based on $Prob_{q}$\;
        $Res_{q} \leftarrow  \text{argTop-k}_{o \in C_i}  ST(q,o) $\;
        {\bf return} $Res_{q}$\;
    }    
\end{algorithm}
\noindent\textbf{Procedures of \Framework.}
The detailed procedures of \Framework are summarized at Algorithm~\ref{alg:workflow}. In the training phase, we train the proposed relevance model and index (lines 2-4). After that, in the indexing phase, we assign all objects to their corresponding clusters as inverted files (lines 6-10). In the query phase, given a new query $q$, we extract its features and then route it to a cluster $C_i$ by the index (lines 13-16). Consequently, we calculate the relevance score between $q$ and each object $o$ in that cluster by the relevance model, and the top-$k$ objects with the highest scores are returned to answer query $q$ (lines 17-18). \revision{In practice, the number of objects assigned to each cluster is  relatively small, and thus we evaluate all objects within the routed clusters
as it is in existing ANNS indexes.}

\noindent\textbf{Complexity Analysis.} Now we analyze the time and space complexity of \Framework. Assuming the dimensionality of the embedding is $d$, the total number of objects is $n$, \revision{the number of clusters is $c$}, and the number of layers in the MLP of the index is $l$. The embeddings $o.emb$ of objects are generated in advance. 


For a given query, the time complexity of \Framework is $O((l-2)d^2+dc+\frac{n}{c}(d+2)+d)$. $O((l-2)d^2+dc)$ is the time complexity for our index (lines 15), which is the inference computation cost of Equation~\ref{eq:L2PO}. $O(\frac{n}{c}(d+2)+d)$ is the time cost of our relevance model (line 17). $\frac{n}{c}$ denotes the number of objects to be checked by the relevance model, which is approximately $\frac{1}{c}$ of the entire dataset. This is because of the even cluster distribution (Verified in Section~\ref{sec:exp-ablation}), which means approximately $\frac{1}{c}$ of the dataset needs to be visited. 
The space complexity of \Framework is $O((l-1)d+dc)+n(d+2)$. $O((l-1)d+dc)$ is the size of the MLP used by our index. $O(nd)$ represents the space required for storing pre-computed $o.emb$, and $O(2n)$ denotes the storage cost for geo-locations.



\section{Experiments}

\label{sec:exp}
In this section, we evaluate the effectiveness and efficiency of our proposed solution for answering \tkskqs by comparing it with state-of-the-art methods on three real-world datasets. We aim to answer the following research questions: 


\begin{itemize}[leftmargin=*,topsep=0pt]
    \item \textbf{RQ1:} \revision{Does our relevance model outperform existing relevance models in terms of effectiveness?}
    \item \textbf{RQ2:} \revision{Does our proposed ANNS index achieve a better effectiveness-efficiency trade-off compared to existing indexes?}
    \item \textbf{RQ3:} \revision{Can our proposed ANNS index be applied to other relevance models designed for embedding based spatial keyword queries?}    
    \item \textbf{RQ4:} How does \Framework scale with the size of dataset?    
    \item \textbf{RQ4:} What are the impacts of our proposed index and different modules in the proposed \Framework?

    
    
\end{itemize}

\begin{table}[!t]
    \centering
    \caption{Datasets Statistics.}
    \label{tab:dataset}
    \renewcommand{\arraystretch}{1.2} 
    \setlength{\arrayrulewidth}{0.5pt}
    \begin{tabularx}{0.5\textwidth}{>{\centering}m{3.6cm}*{2}{X}m{1.5cm}}
    \toprule[1pt]
    \multirow{2}{*}{} & Beijing & Shanghai & Geo-Glue\\
    \midrule[1pt]
    Number of Pois & 122,420 & 116,859 & 2,849,754\\
    Number of Queries & 168,998 & 127,183 & 90,000\\
    Number of Records & 233,343 & 182,634 & 90,000\\
    \midrule
    Number of Train Queries & 136,890 & 103,019 & 50,000 \\
    Number of Val Queries& 15,209 & 11,446 & 20,000 \\
    Number of Test Queries& 16,899 & 12,718 & 20,000 \\
    \midrule
    Number of Train Records & 189,027 & 148,017 & 50,000 \\
    Number of Val Records& 21,034 & 16,492 & 20,000 \\
    Number of Test Records& 23,282 & 18,125 & 20,000\\    
    \bottomrule[1pt]
    \end{tabularx}
\end{table}

\subsection{Experimental Setup}
\label{sec:exp-setup}
\noindent\textbf{Datasets.} To evaluate the effectiveness and efficiency of our proposed retriever \Framework, we utilize three benchmark datasets: Beijing, Shanghai, and Geo-Glue. Among them, the Beijing and Shanghai datasets~\cite{liu2023effectiveness} are provided 
a Chinese retail services platform. Users submit a query through the platform, which consists of a query location and a set of keywords. Subsequently, the Points of Interest (POIs) that users clicked on are recorded in the query log, and is considered as a ground-truth. Note that the click-through data recorded in the search log may be the only feasible way to get a large scale of ground truth data for spatial keyword queries~\cite{liu2023effectiveness}. The explicit feedback such as ratings is very challenging to collect~\cite{DBLP:conf/www/HeLZNHC17}. Note that using query logs is also the popular way of generating ground truth in the Information Retrieval literature~\cite{DBLP:conf/www/HeLZNHC17,DBLP:conf/cikm/HuangHGDAH13,DBLP:conf/www/YaoTCYXD021,DBLP:conf/kdd/Joachims02}. Therefore, following the previous work~\cite{liu2023effectiveness}, we utilize the two datasets to evaluate our proposed solution \Framework and treat the clicked POIs as ground truth relevant objects to the corresponding queries. In the Geo-Glue dataset~\cite{li2023geoglue,DBLP:conf/sigir/DingCXHLZX23}.
the POIs  are crawled from OpenStreetMap\footnote{https://www.openstreetmap.org/}, and the queries and the corresponding ground truth POIs are manually generated by domain experts. Notably, in the released Geo-Glue dataset, the coordinates of objects and queries are modified due to privacy considerations, which results in many objects with identical geo-locations.

The release of query log datasets from the industry is highly restricted. {As a result, to the best of our knowledge, there are no other public datasets that contain ground-truth query results or query logs, except for the three datasets used in our experiments.} To investigate the scalability of the proposed framework, we conduct a scalability study to show \Framework's efficiency on larger datasets, where the Geo-Glue dataset is augmented with more crawled POIs from OpenStreetMap (as to be shown in Section~\ref{sec:exp-scalibility})

\begin{table}[!t]
    \centering
    \caption{\revision{Hyperparameter settings on three datasets.}}
    \label{tab:hyper}
    \renewcommand{\arraystretch}{1.2} 
    \setlength{\arrayrulewidth}{0.5pt}
    \small
    \begin{tabularx}{0.5\textwidth}{>{\centering}m{3.6cm}*{2}{X}m{1.5cm}}
    \toprule[1pt]
    Hyperparameters & Beijing & Shanghai & Geo-Glue\\
    \midrule[1pt]
    spatial footstep $t$ & 1000 & 1000 & 1000 \\
    pseudo-label's $\text{neg}_\text{start}$ & 50,000 & 60,000 & 180,000\\
    pseudo-label's $\text{neg}_\text{end}$ & 55,000 & 65,000 & 181,000\\
    number of cluster $c$ & 20 & 20  & 300 \\  
    \bottomrule[1pt]
    \end{tabularx}
\end{table}

\noindent\textbf{Dataset Split.} The statistics of datasets are stated in Table~\ref{tab:dataset}, where each record represents a single ground-truth label between an object and a query, and each query may have multiple ground-truth objects. 
For the Beijing and Shanghai datasets, to ensure a fair comparison, we follow the previous split strategy~\cite{liu2023effectiveness}, where 90\% of queries and their corresponding ground-truth records are used as the training set and the remaining queries as the test set. From the training set, we randomly choose 10\% data as the validation set to tune hyperparameters. For the Geo-Glue dataset~\cite{DBLP:conf/sigir/DingCXHLZX23}, we follow the provided splits for training, validation, and test data.
\begin{table*}[t]
    \centering
    \caption{Comparison of relevance models across three datasets by brute-force search.}
    \label{tab:relevance_model}
    \renewcommand{\arraystretch}{1.5} 
    \setlength{\arrayrulewidth}{0.5pt}
    \begin{tabularx}{1\textwidth}{>{\centering}m{2cm}*{4}{X}*{4}{X}*{4}{X}}
        \toprule[1pt]
        \multirow{2}{*}{} & \multicolumn{4}{c}{\textbf{Beijing}} & \multicolumn{4}{c} {\textbf{Shanghai}} & \multicolumn{4}{c} {\textbf{Geo-Glue}} \\
        \cmidrule(lr){2-5} \cmidrule(lr){6-9} \cmidrule(lr){10-13} 
        & \multicolumn{2}{c}{\textbf{Recall}} & \multicolumn{2}{c}{\textbf{NDCG}} & \multicolumn{2}{c}{\textbf{Recall}} & \multicolumn{2}{c}{\textbf{NDCG}}  &
        \multicolumn{2}{c}{\textbf{Recall}} & \multicolumn{2}{c}{\textbf{NDCG}} \\
        \cmidrule(lr){2-3} \cmidrule(lr){4-5} \cmidrule(lr){6-7} \cmidrule(lr){8-9} 
        \cmidrule(lr){10-11} \cmidrule(lr){12-13}
        & \textbf{@20} & \textbf{@10} & \textbf{@5} & \textbf{@1} & \textbf{@20} & \textbf{@10} & \textbf{@5} & \textbf{@1} & \textbf{@20} & \textbf{@10} & \textbf{@5} & \textbf{@1} \\
        \midrule[1pt]
        TkQ  & 0.5740 & 0.5283 & 0.4111 & 0.3302 & {0.6746} & 0.6380 & 0.5044 & 0.4009 & 0.5423 & \underline{0.5023} & \underline{0.3847} & 0.3051 \\

        PALM & 0.3514 & 0.3098 & 0.2077 & 0.1343 & 0.4617 & 0.4023 & 0.2065 & 0.1223 &  N/A & N/A &  N/A & N/A   \\        

        DrW & \underline{0.6968} & \underline{0.6316} & \underline{0.4814} & \underline{0.3791} & \underline{0.7689} & \underline{0.7159} & \underline{0.5394} & \underline{0.4114} & N/A & N/A &  N/A & N/A \\
        
   

        
        MGeo& N/A & N/A & N/A & N/A & N/A & N/A & N/A & N/A & \underline{0.7049} & N/A & N/A  & \underline{0.5270} \\ 
        \midrule
        \textbf{\Framework-R} & \textbf{0.8156} & \textbf{0.7545} & \textbf{0.5913} & \textbf{0.4989} & \textbf{0.8361}  & \textbf{0.7924} &\textbf{0.6445} & \textbf{0.5397} & \textbf{0.8393} & \textbf{0.8033} & \textbf{0.6837} & \textbf{0.5887} \\           
        (Gain) & 17.04\% & 19.45\% & 22.82\% & 31.60\% & 10.63\% & 19.48\% & 19.06\% & 18.14\% & 19.06\% & 59.92\% & 77.72\% & 11.70\% \\ 
        \bottomrule[1pt]
    \end{tabularx}
\end{table*}


\noindent\textbf{Effectiveness metric.} Following previous studies~\cite{DBLP:conf/sigir/DingCXHLZX23,liu2023effectiveness}, we use two 
metrics, i.e., the Recall and Normalized Discounted Cumulative Gain (NDCG), to evaluate the effectiveness.
Recall@$k$ evaluates the proportion of positive objects contained in the top-$k$ candidates for a given query. NDCG@$k$ considers the order of ground-truth objects in the retrieved objects, reflecting the quality of the ranking in the retrieved list. We assign the graded relevance of the result at position $i$ as $rel_{i} \in \{0,1\}$, where $rel_{i}=1$ when the object is relevant to the query, otherwise $rel_{i}=0$. More details can be found in \cite{liu2023effectiveness}. Specifically, the following metrics are utilized: Recall@10, Recall@20, NDCG@1, and NDCG@5. 


\noindent\textbf{Relevance Model Baselines.} We compare our 
relevance model, denoted as \Framework-R, with existing state-of-the-art spatio-textual relevance models. 
\begin{itemize}[leftmargin=*,topsep=0pt]
\item TkQ~\cite{congEfficientRetrievalTopk2009}: It uses traditional relevance model BM25~\cite{DBLP:journals/ftir/RobertsonZ09} to evaluate text relevance and treats $1-SDist(q.loc,o.loc)$ as spatial relevance. The weight parameter $\alpha$ is manually tuned from 0 to 1 with a footstep of 0.1, and the best effectiveness is achieved when $\alpha=0.4$ for all three datasets. 
\item PALM~\cite{DBLP:conf/aaai/ZhaoPWCYZMCYQ19}: This method employs deep neural networks for query-object spatio-textual relevance.
\item DrW~\cite{liu2023effectiveness}: This is the 
newest deep relevance based method for answering \tkskqs. 
\item MGeo~\cite{DBLP:conf/sigir/DingCXHLZX23}: This is a recent deep learning based method for answering \tkskqs. Note that we cannot reproduce the experimental results reported in the paper by running the official code, and thus we use the evaluation results from the original paper.
\end{itemize}

\noindent\revision{\textbf{Index Baselines.} We compare our proposed index with existing state-of-the-art indexes. 
Specifically, we select the IR-Tree~\cite{congEfficientRetrievalTopk2009} to accelerate the TkQ search, serving as an index baseline. As discussed in Section~\ref{sec:related}, existing ANNS indexes fall into two categories and our method belongs to the first category. We select the state-of-the-art ANNS indexes in the first category, i.e., IVF~\cite{jegouProductQuantizationNearest2011}, LSH~\cite{johnson2019billion}, and HNSW~\cite{malkovEfficientRobustApproximate2020}, as index baselines. For the second category of methods, such as product quantization based methods, their standalone efficiency is no better than a brute-force search\footnote{https://github.com/facebookresearch/faiss/issues/148}. Therefore, following previous work~\cite{jegouProductQuantizationNearest2011}, we choose IVFPQ as an index baseline, which integrates PQ with the IVF index from the first category and is considered as a state-of-the-art solution. As discussed  in Section~\ref{sec:motivations}, existing ANNS indexes are limited to considering only textual embeddings, neglecting the spatial factor. To mitigate this limitation, 
we extend the IVF index to include both embedding and spatial factors
, which is denoted as IVF$_S$ index, as an index baseline. Similar to \Framework, which uses a retrieve and rerank pipeline, we utilize the aforementioned indexes to retrieve a subset of objects and then rerank the retrieved objects with our relevance model \Framework-R for a fair comparison. The integration of our proposed index and \Framework-R is \Framework, and the other methods are detailed below.}
\begin{itemize}[leftmargin=*,topsep=0pt]
\item \revision{TkQ+\Framework-R}: This method employs TkQ to retrieve top-$k$ objects and 
reranks these objects by \Framework-R.
\item \revision{IVF+\Framework-R}: This method constructs an IVF index over the embeddings produced by \Framework-R, and the objects within the selected index cluster are reranked by \Framework-R. Notably, this method requires two parameters: the number of clusters $c$, and the number of clusters to route for each query and object $cr$. It does not involve $k$. We set $c$ and $cr$ to be the same as our method across three datasets for a fair comparison.

\item \revision{IVF$_S$+\Framework-R}: This method follows the IVF+\Framework-R pipeline but diverges in the clustering approach. Instead of only applying k-means to text embeddings, this approach utilizes k-means on the weighted sum of both geo-location and text embedding factors. The weight $\alpha$ is manually tuned from 0 to 1 with a footstep of 0.1, and the best effectiveness is achieved when $\alpha=0.9$ for all these datasets.

\item  \revision{LSH+\Framework-R}: This method constructs an LSH index with the embeddings produced by \Framework-R, which retrieves the top-$k$ relevant objects by fetching similar embeddings in the same buckets, and reranks the retrieved objects by \Framework-R. Following previous work~\cite{johnson2019billion}, we set the length of the hash code $nbits$ to 128.

\item  \revision{HNSW+\Framework-R}: This method constructs an HNSW index over the embeddings produced by \Framework-R, which retrieves the top-$k$ relevant objects by conducting beam searches over the proximity graph, and reranks the retrieved objects by \Framework-R. Following previous work~\cite{malkovEfficientRobustApproximate2020}, we set the number of links $M$ to 48 and $efConstruction$ to 100. 

\item  \revision{IVFPQ+\Framework-R}: This method integrates the IVF index with the product quantization technique~\cite{jegouProductQuantizationNearest2011} to retrieve objects. The retrieved objects are reranked by \Framework-R. We set the number of clusters $c$ to be the same as our index. Following the instruction~\cite{johnson2019billion}, we set the number of centroids $w$ to 32 and the number of bits $nbits$ to 8, the number of clusters to search $cr$ at 2. 
Here, we set a larger value for $cr$ than that in our index since the product quantization accelerates computations within clusters, allowing access to more clusters with comparable time costs. Notably, the maximal $k$ supported by the Faiss library is 2,048, therefore we set $k$ to 2,048 across three datasets. 
\end{itemize}
An important hyperparameter is $k$, which represents the number of objects retrieved by the indexes and reranks by LIST-R. To ensure a fair comparison, we set $k$ to 5,000 for the baselines that involve $k$ on the Beijing and Shanghai datasets, which is the average number of objects retrieved by our index per query. On the geo-glue dataset, $k$ is set to be 30,000 for the same reason above. \revision{To investigate the capability of \Framework to trade off effectiveness and efficiency, we evaluate the effect of the hyperparameter $cr$ and $k$ in Section~\ref{sec:exp-effectiveness-efficiency-tradeoff}.}

\noindent\textbf{Implementations.} The relevance model and the index are trained using Pytorch. During the query phase, the index and all relevance models are inferred in C++ by the ONNX system\footnote{https://github.com/onnx/onnx}. In our proposed relevance model, we utilize the bert-base-Chinese pre-trained model\footnote{https://huggingface.co/bert-base-chinese} from the huggingface Library~\cite{wolf2019huggingface} as encoders, which is the same as previous work~\cite{liu2023effectiveness,DBLP:conf/sigir/DingCXHLZX23}. \revision{The hyperparameter settings of \Framework are detailed in Table~\ref{tab:hyper}.} Specifically, the spatial footstep controller $t$ of $\mathcal{T}$ is set to 1000 for all three datasets. The hyperparameters to control the generation of pseudo labels, $\text{neg}_\text{start}$ and $\text{neg}_\text{end}$, are set to 50,000 and 55,000 for the Beijing dataset, 60,000 and 65,000 for the Shanghai dataset, and 180,000 and 181,000 for the Geo-Glue dataset.
We empirically set the cluster number $c$ to approximately $\frac{n}{10,000}$, i.e.,  20 for the Beijing and Shanghai datasets and 300 for the Geo-Glue dataset. This is because ranking 10,000 objects by the relevance model is computationally feasible and does not notably compromise effectiveness. The number of clusters to route $cr$ for queries and objects is set to 1 by default, with different $cr$ settings shown in Section~\ref{sec:exp-effectiveness-efficiency-tradeoff}. The implementations of DrW are from publicly available source codes, and we make use of the implementation of IVF, LSH, IVFPQ, and HNSW provided by the Faiss library~\cite{johnson2019billion}, while others are implemented by ourselves. The Faiss library is implemented in C++, providing a fair comparison with our index. 
Our default experiment environment is CPU: Intel(R) Core(TM) i9-10900X CPU@3.70GHz, Memory: 128G, and GPU: V100 32GB.

\begin{figure*}[!t]
\centering
\subcaptionbox{Beijing\label{fig:exp-beijing}}{
\includegraphics[width=0.9\textwidth]{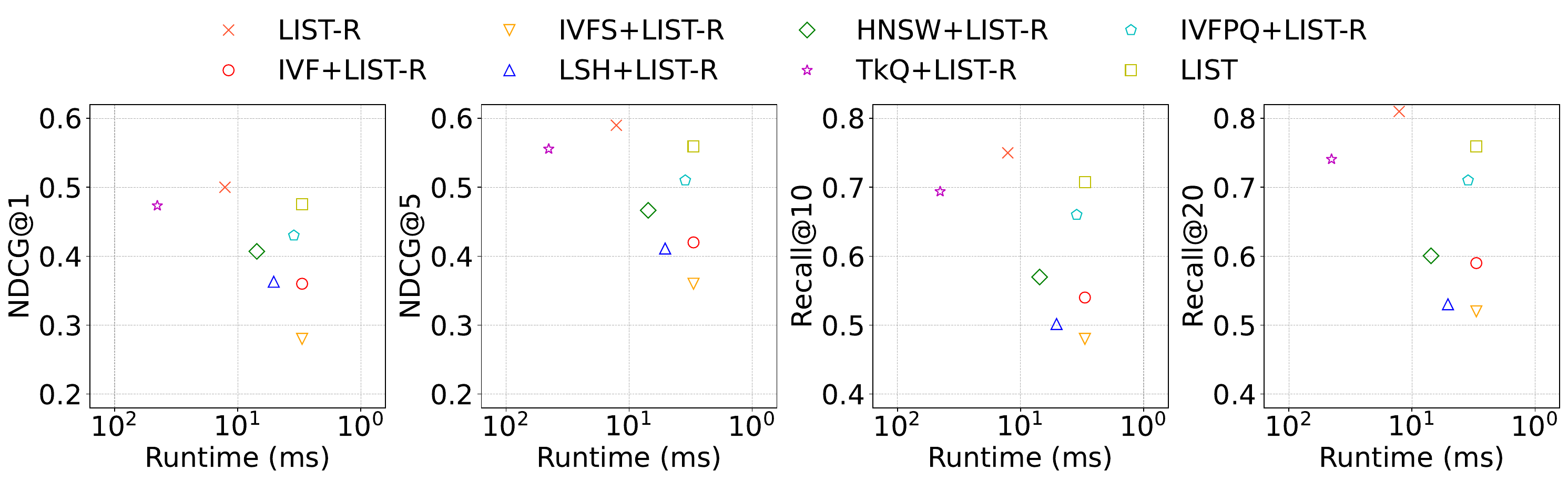}
}
\subcaptionbox{Shanghai\label{fig:exp-shanghai}}{
\includegraphics[width=0.9\textwidth]{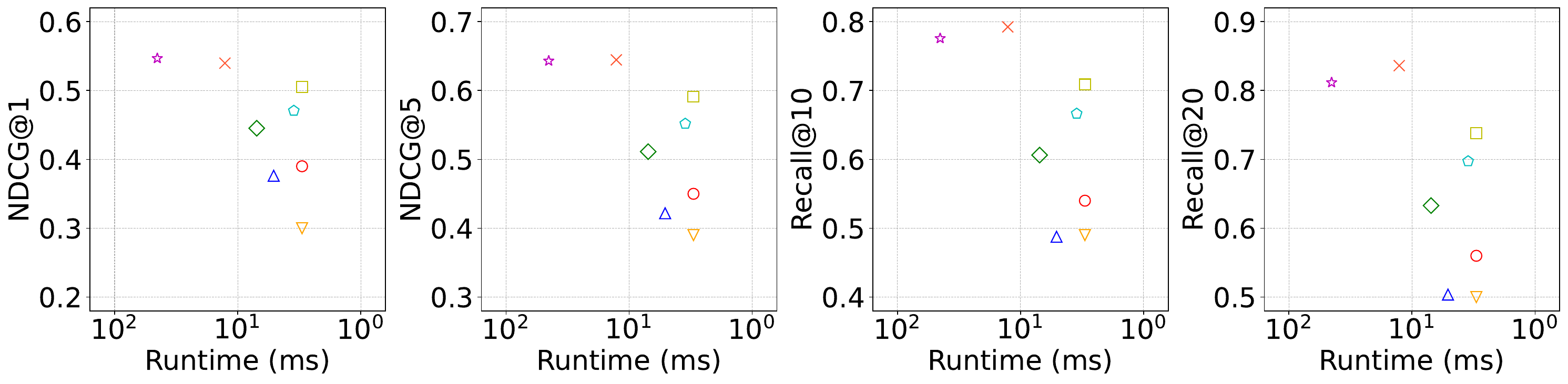}
}
\subcaptionbox{Geo-Glue\label{fig:exp-geoglue}}{
\includegraphics[width=0.9\textwidth]{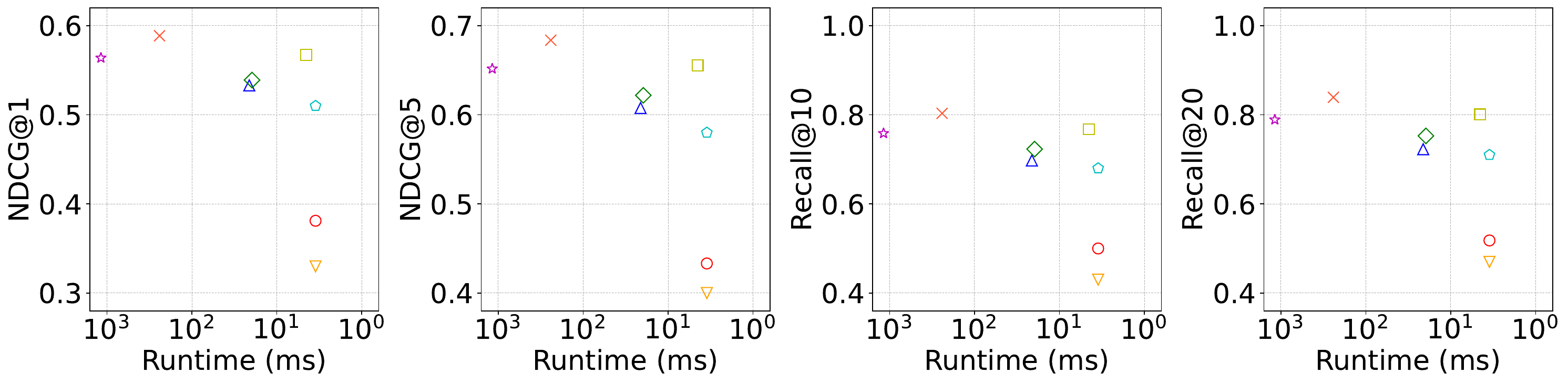}
}
\caption{\revision{The effectiveness-efficiency trade-off results (upper and right is better).}}
\label{fig:exp-effective-efficiency}
\end{figure*}

\subsection{Effectiveness of Proposed Relevance Model (RQ1)}
\label{sec:exp-effectiveness}
\noindent\textbf{Effectiveness of Relevance Model \Framework-R.} To validate our proposed relevance model's effectiveness, denoted as \Framework-R, we compare it with other relevance models across the three datasets. All relevance models perform a brute-force search over the entire dataset to identify the top-k objects. Table~\ref{tab:relevance_model} reports the effectiveness of the evaluated methods. DrW and PLAM cannot be evaluated on the Geo-Glue dataset via brute-force search because of their slow querying speed, requiring more than a day for evaluation. We have the following findings: (1) \Framework-R consistently outperforms all the baseline models on all three datasets across every metric. Specifically, \Framework-R achieves up to a 31.60\% improvement over the best baseline on NDCG@1 and up to a 59.92\% improvement on recall@10. (2) Traditional ranking methods are less effective than deep relevance models. On the three datasets, TkQ is outperformed by DrW. 
This can be attributed to the word mismatch issue discussed in Section~\ref{sec:motivations}. Notably, PALM is outperformed by TkQ, which is consistent with the results reported by~\cite{liu2023effectiveness}. 

\begin{figure*}[!t]
\centering
\subcaptionbox{Beijing\label{fig:exp-beijing-k}}{
\includegraphics[width=0.9\textwidth]{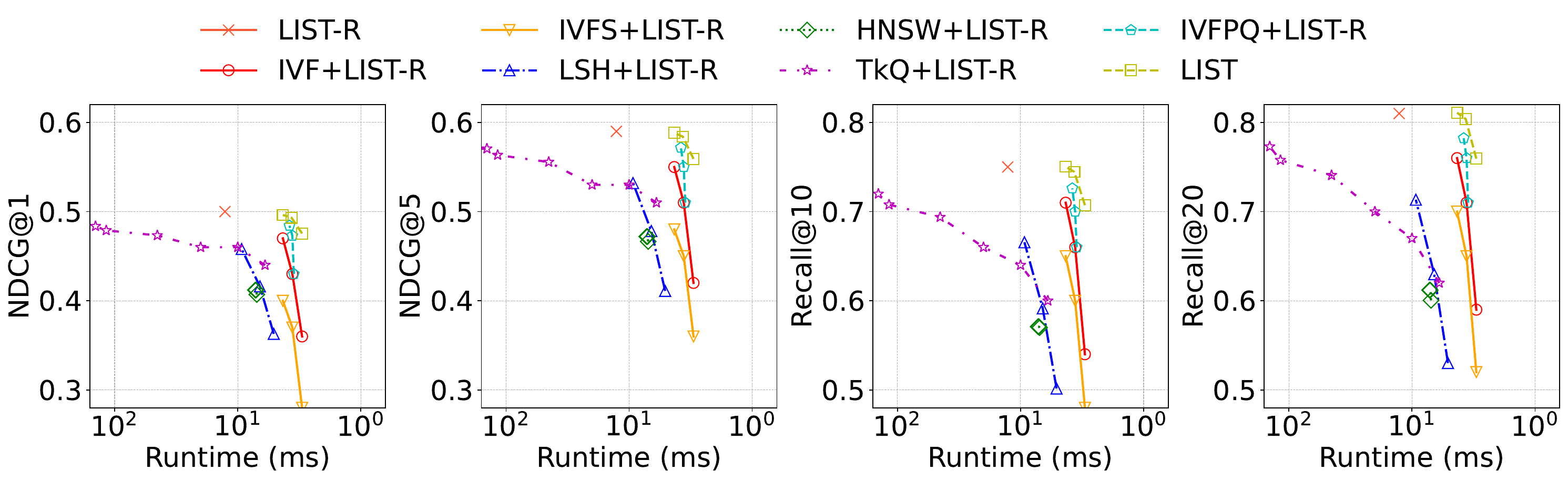}
}
\subcaptionbox{Shanghai\label{fig:exp-shanghai-k}}{
\includegraphics[width=0.9\textwidth]{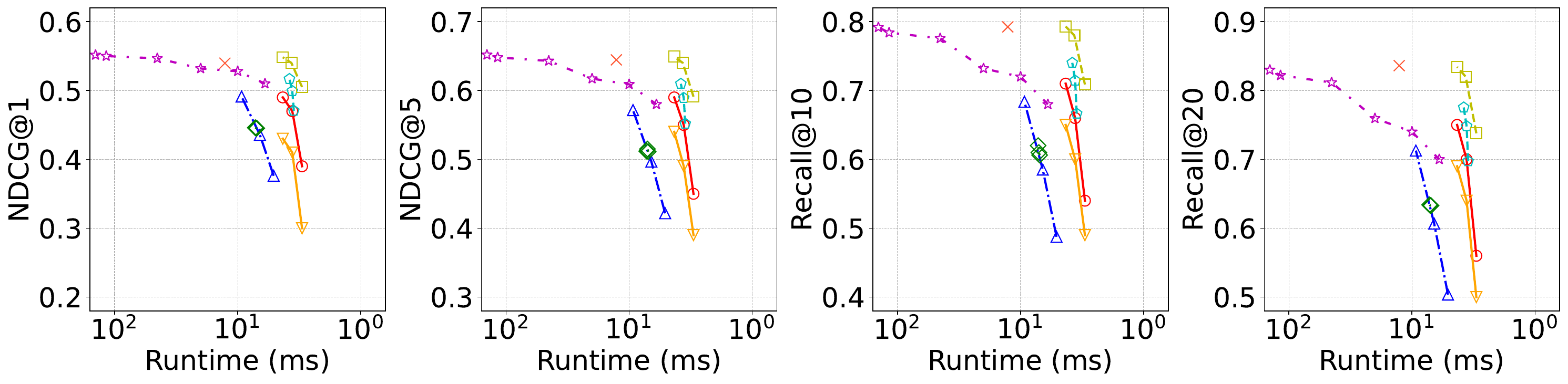}
}
\caption{\revision{The Effectiveness-Speed trade-off curve varies with the number of objects retrieved (top-$k$) and the number of clusters to route ($cr$) (up and right is better).}}
\label{fig:exp-effective-efficiency-tradeoff}
\end{figure*}

\subsection{\revision{Effectiveness-Efficiency Trade-off of Proposed Index (RQ2)}}
\label{sec:exp-effectiveness-efficiency-tradeoff}

\noindent\revision{\textbf{Effectiveness-efficiency trade-off of \Framework.} We investigate the effectiveness-efficiency trade-off of \Framework by comparing it with the state-of-the-art index baselines, which all use our relevance model. The evaluation results are shown in Figure~\ref{fig:exp-effective-efficiency}, where LIST-R represents the brute-force search using our relevance model. We have the following findings: (1) Overall, \Framework consistently outperforms all baselines, offering the best trade-off between effectiveness and efficiency across all three datasets. Specifically, compared to 
LIST-R, \Framework is an order of magnitude faster on the Beijing and Shanghai datasets and three orders of magnitude faster on the Geo-Glue dataset. While most ANNS indexes have similar runtime to \Framework, their effectiveness falls short. 
(2) Directly applying ANNS indexes to spatial keyword queries results in a significant drop in effectiveness. Compared to LIST-R, which uses brute-force search, existing ANNS indexes sacrifice considerable effectiveness for improved efficiency. This could be attributed to the fact that these ANNS indexes do not consider the spatial factor. 
(3) When TkQ is used as an index baseline, it 
is significantly slower. The reason is that the IR-Tree is not designed to retrieve a large number of objects.  
As the 
hyperparameter $k$ increases, the IR-Tree's pruning ability declines, which is consistent with previous empirical results~\cite{congEfficientRetrievalTopk2009}. (4) Simply modifying the existing ANNS index to incorporate the spatial factor is even less effective. IVF+LIST-R consistently outperforms IVF$_S$+LIST-R on the three datasets. This confirms that manually assigning a weight to balance the two factors is 
ineffective. }

\noindent\textbf{Trade-off Study by varying the number of objects retrieved (top-$k$) and the number of clusters to route ($cr$).} \revision{We examine \Framework's ability to trade off effectiveness and efficiency. 
Here, we vary the hyperparameter $cr$, which represents the number of clusters to route, from 1, 2, to 3 on the Beijing and Shanghai datasets. For baselines that share the same hyperparameter with \Framework, i.e., IVF+\Framework-R, IVF$_S$+\Framework-R, and IVFPQ+\Framework-R, we maintain the same increment for $cr$. For other methods, we adjust the hyperparameter $k$, which represents the number of objects retrieved by the indexes, to be close to the number of objects retrieved by our proposed index, adjusting it from 5,000, 10,000, to 20,000 on the Beijing and Shanghai datasets. For TkQ+\Framework-R, due to its slow query processing speed, we adjust $k$ from 100, 500, 1,000, 5,000, and 10,000 to 20,000. We present the trade-off results in Figure~\ref{fig:exp-effective-efficiency-tradeoff}. We have the following findings: (1) \Framework consistently outperforms other methods, providing a better trade-off between effectiveness and efficiency. (2) Compared to other methods, the effectiveness of TkQ+LIST-R increases more slowly as $k$ increases, while the time overhead rises significantly. This is due to the word mismatch issue discussed in Section~\ref{sec:motivations}. Many relevant objects without word overlap with the query remain hard to retrieve as $k$ increases. (3) HNSW+LIST-R 
is almost not affected by the number of retrieved objects. 
This is because HNSW builds a sparse proximity graph, limiting the reachable objects for each query. Therefore, when $k$ exceeds the number of reachable objects, its performance remains unchanged.}

\noindent\textbf{Memory Consumption of \Framework.} 
The memory consumption of LIST is composed of three parts: the memory used by the proposed relevance model, the memory used by the proposed index, and the memory used for object text embeddings that are produced in advance. The experiment results are presented in Table~\ref{tab:memory}, which demonstrates the remarkable memory efficiency of \Framework. Compared with LSH+LIST-R, TkQ+LIST-R, and HNSW+LIST-R, \Framework requires less memory. The reason is that our index stores only a lightweight MLP $c$-cluster classifier. This storage requirement is less than the memory consumption of the proximity graph of HNSW, hash tables of LSH, and the inverted file of the IR-Tree.
\begin{table}[h]
    \centering
     \caption{Memory usage on three datasets (MB).}
    \label{tab:memory}
    \renewcommand{\arraystretch}{1.5} 
    \setlength{\arrayrulewidth}{0.5pt}
    \small
    \begin{tabularx}{0.5\textwidth}{>{\centering}m{2.5cm}*{3}{X}}
    \toprule[1pt]
    \multirow{2}{*}{} & Beijing & Shanghai & Geo-Glue\\
    \midrule[1pt]
    TkQ+LIST-R & 719 & 857 & 12,037\\
    IVF+LIST-R & 508 & 505 & 8,510 \\  
    LSH+LIST-R & 513 & 511 & 8,638 \\ 
    HNSW+LIST-R & 548 & 545 & 9,427 \\ 
    IVFPQ+LIST-R & 508 & 505 & 8,520 \\  
    \midrule
    \textbf{\Framework} & 508 & 505 & 8,515 \\    
    \bottomrule[1pt]
  \end{tabularx}
\end{table}

\begin{figure*}[!t]
\centering
\subcaptionbox{Beijing\label{fig:exp-robeta}}{
\includegraphics[width=0.9\textwidth]{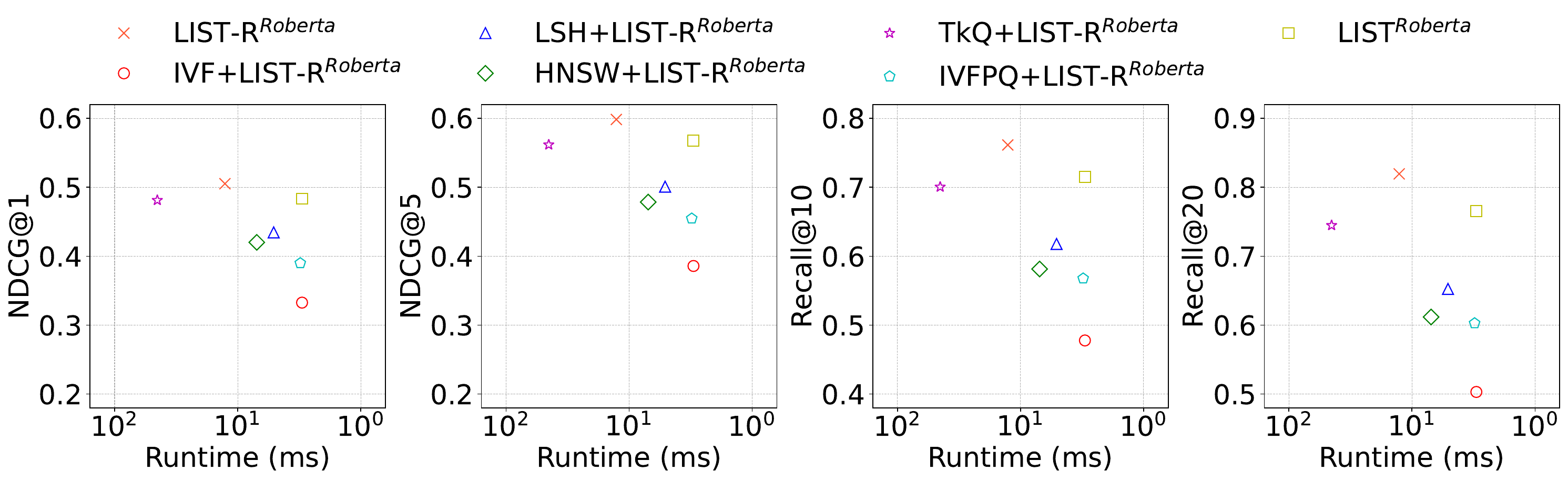}
}
\subcaptionbox{Beijing\label{fig:exp-linear}}{
\includegraphics[width=0.9\textwidth]{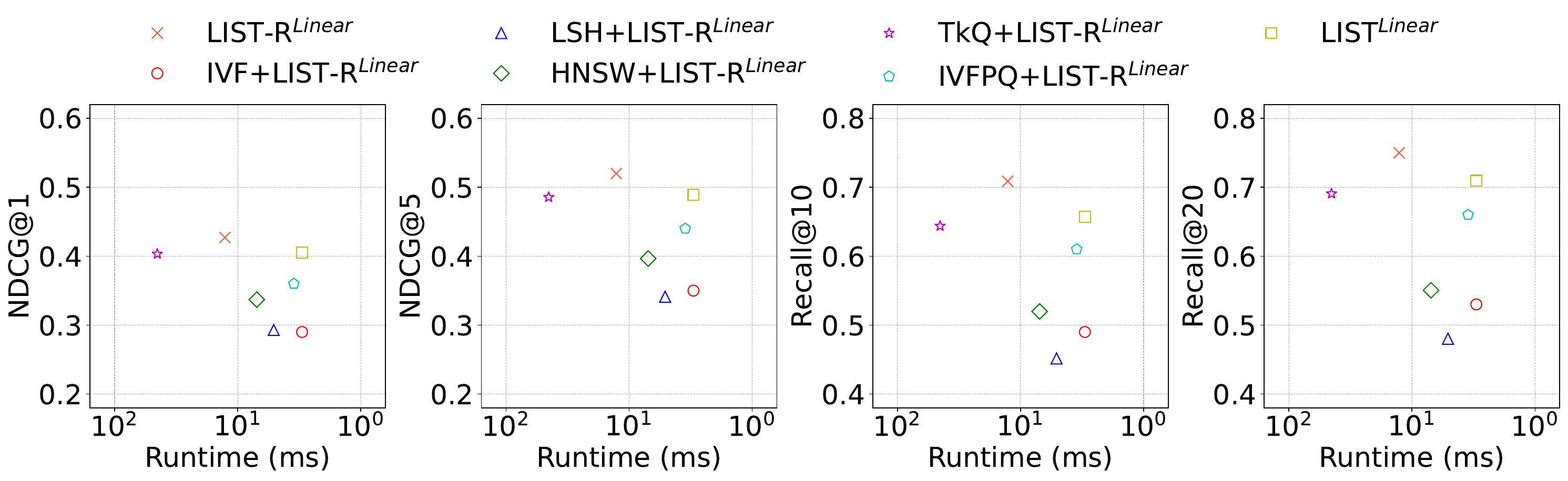}
}
\caption{\revision{The effectiveness-efficiency trade-off results for different relevance models (upper and right is better).}}
\label{fig:exp-generalization}
\end{figure*}

\subsection{\revision{Generalization Study of the Proposed Index (RQ3)}}
\noindent\revision{\textbf{Generalization Study.} We explore the applicability of our proposed index to other relevance models designed for embedding based spatial keyword queries. Specifically, we introduce two variants of our relevance model: LIST-R$^{Roberta}$ and LIST-R$^{Linear}$. Specifically, in LIST-R$^{Roberta}$, the BERT model in the dual-encoder module is replaced by another pre-trained language model RoBERTa\footnote{\url{https://huggingface.co/clue/roberta_chinese_base}}~\cite{wolf2019huggingface}. In LIST-R$^{Linear}$, the learned spatial relevance module is replaced with the linear function of distance. We train the two baselines on the Beijing dataset and use them to generate embeddings in advance. We then use our proposed index to build an index on these embeddings and compare it with existing index baselines. The comparison of our proposed index with the baseline indexes is shown in Figure~\ref{fig:exp-generalization}. The experimental results demonstrate that our proposed index maintains a similar advantage over baseline indexes when applied to other relevance models. 
}

\subsection{Scalibility Study (RQ4)}
\label{sec:exp-scalibility}
\begin{figure}[!hbtp]
\centering
\subcaptionbox{\Framework-R\label{fig:scalibilty-LISR-R}}{
\includegraphics[width=0.45\columnwidth]{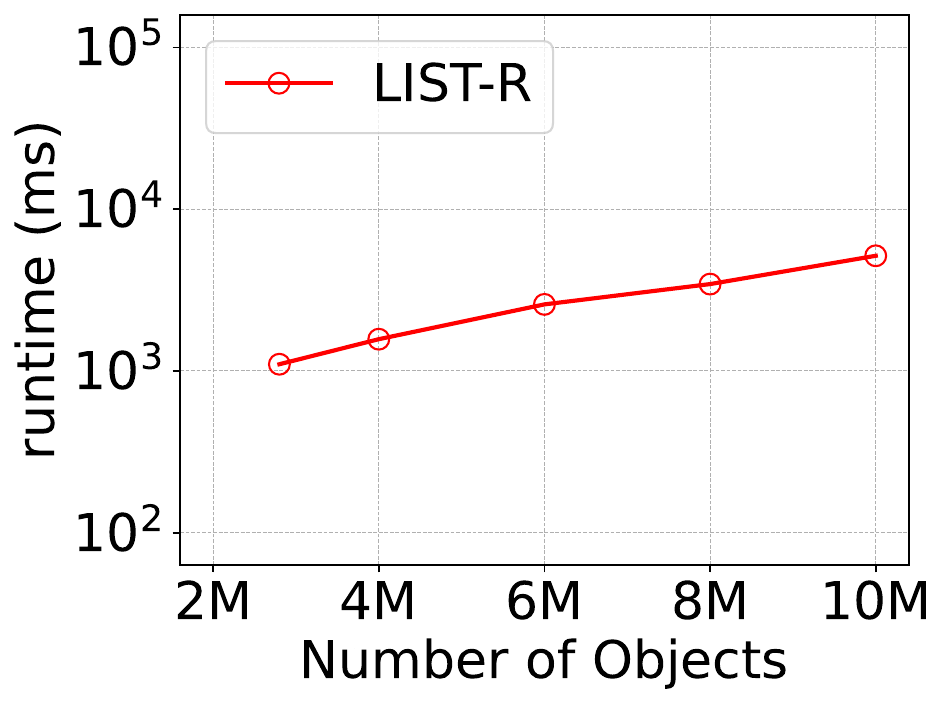}
}
\subcaptionbox{\Framework\label{fig:scalibilty-LIST}}{
\includegraphics[width=0.45\columnwidth]{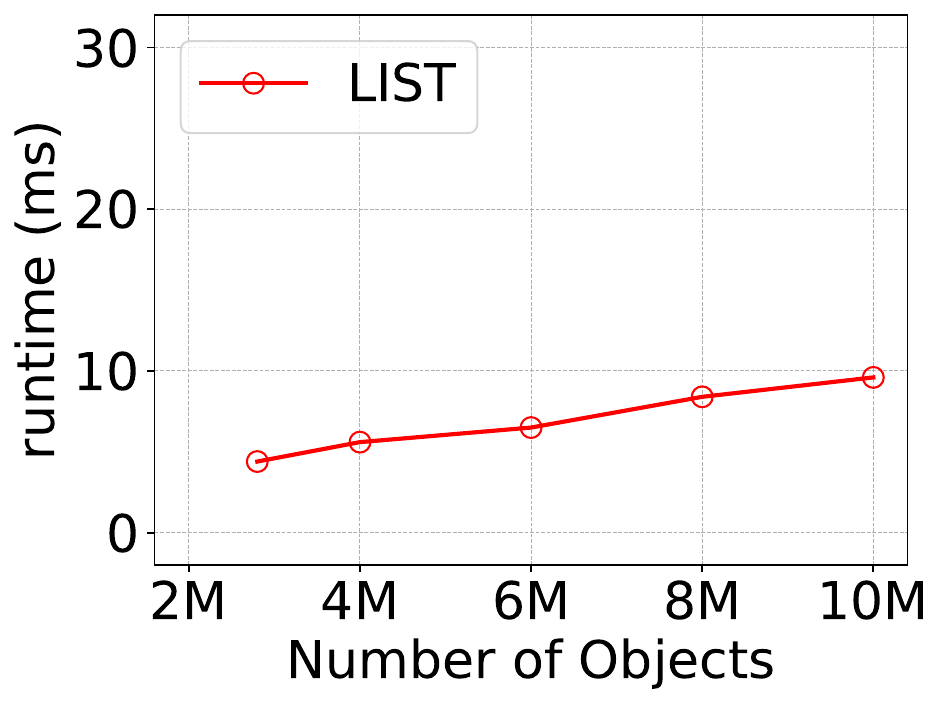}
}
\caption{Scalability study on the Geo-Glue dataset.}
\label{fig:scalibility}
\end{figure}

\noindent\textbf{Scalability Study.} We evaluate the scalability of \Framework and the proposed relevance model \Framework-R. We supplement the Geo-Glue dataset with new POIs crawled from Open Street Maps in Hangzhou. Since there is no ground-truth data for the relevance between the newly crawled POIs and the existing queries, we only report the efficiency of our proposed retriever \Framework and our relevance model \Framework-R. For the newly acquired POIs, we utilize the trained dual-encoder module and index to partition them into distinct clusters. We report the runtime of \Framework and \Framework-R on Figure~\ref{fig:scalibility}.
We observe that, as the number of objects increases, the runtime of \Framework and \Framework-R scales 
linearly. 





\subsection{Ablation Studies (RQ4)}
\label{sec:exp-ablation}

\noindent\textbf{Cluster Quality Study.} To illustrate the quality of produced clusters by our index, we conduct a comparison study. We use the proposed relevance model to generate embeddings and then employ our index (denoted as \Framework-I) and IVF index to produce clusters separately. We present the cluster results of $IF(C)$ and $P(C)$ in Table~\ref{tab:index_ablation_efficiency}, which shows that our index achieves much higher precision and obtains comparative imbalance factors compared with IVF index. 





\begin{figure}[h]
\begin{center}
\begin{subfigure}[b]{0.45\columnwidth}
    \centering
    \includegraphics[width=\textwidth]{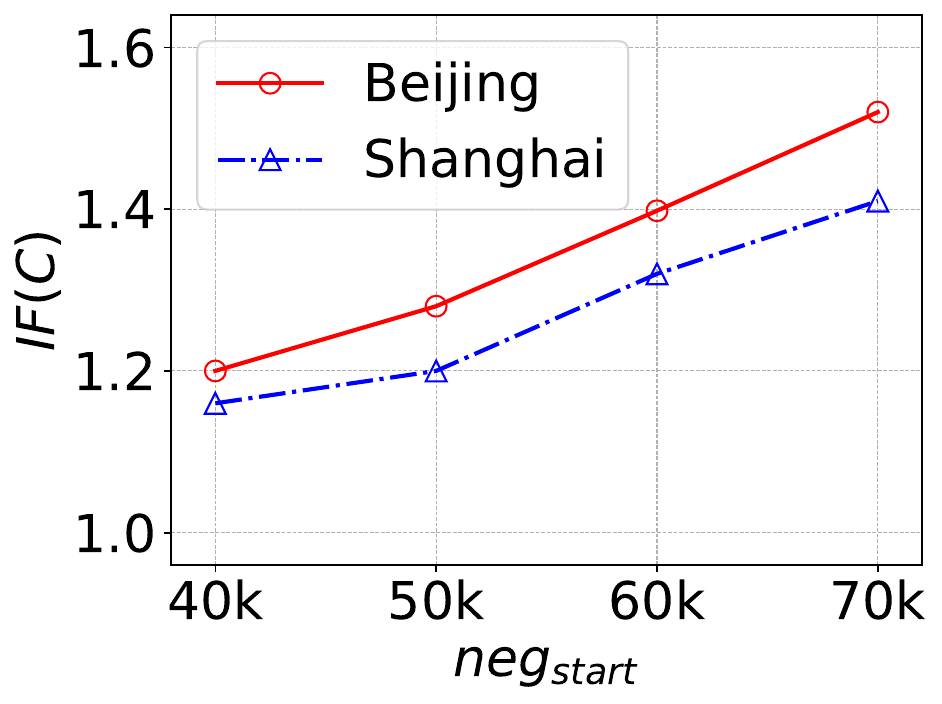}
    \caption{The impact over $IF(C)$}
    \label{fig:negstart-if}
\end{subfigure}
\hfill
\begin{subfigure}[b]{0.45\columnwidth}
    \centering
    \includegraphics[width=\textwidth]{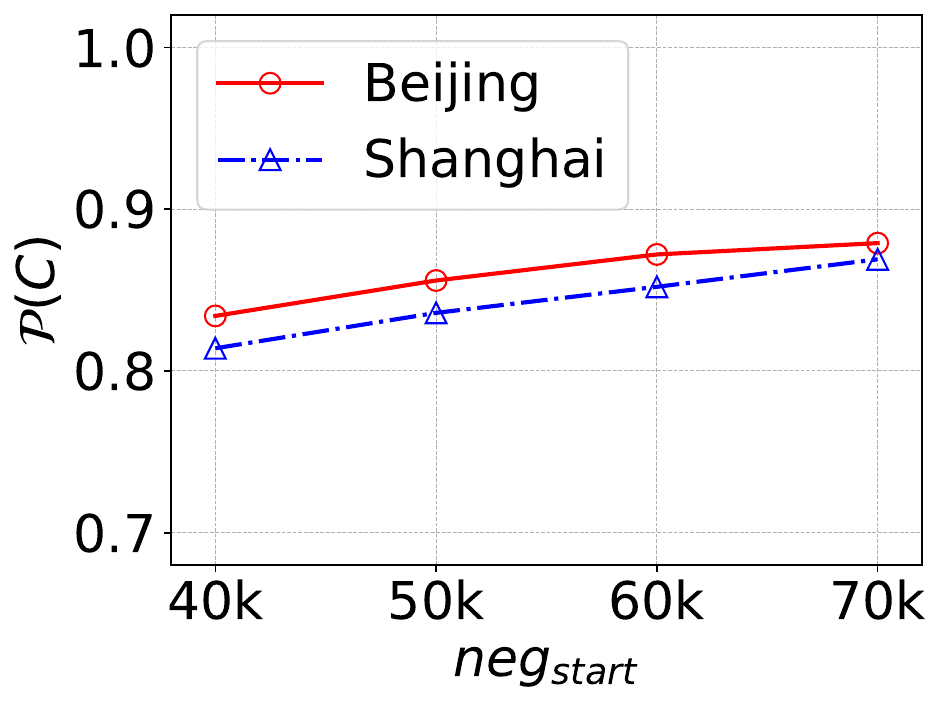}
    \caption{The impact over $P(C)$}
    \label{fig:negstart-pre}
\end{subfigure}
\caption{The impact of neg$_\text{start}$ over the cluster quality.}
\label{fig:negstart}
\end{center}
\end{figure}

\begin{table}[h]
    \centering
    \caption{Comparison of the quality of clusters.}
    \label{tab:index_ablation_efficiency}
    \renewcommand{\arraystretch}{1.5} 
    \setlength{\arrayrulewidth}{0.45pt}
    \begin{tabularx}{0.5\textwidth}{>{\centering}m{2cm}*{2}{X}*{2}{X}}
        \toprule[1pt]
        \multirow{2}{*}{} & \multicolumn{2}{c}{\textbf{Beijing}} & \multicolumn{2}{c} {\textbf{Shanghai}} \\
        \cmidrule(lr){2-3} \cmidrule(lr){4-5} 
        & \textbf{\text{IF}(C)} & \textbf{$P$(C)} & \textbf{\text{IF}(C)} & \textbf{$P$(C)} \\
        \midrule[1pt]
        IVF & 1.31 & 0.6774 & 1.33 & 0.6418 \\     
        \Framework-I & 1.49 & 0.8907 & 1.43 & 0.8382 \\
        \bottomrule[1pt]
    \end{tabularx}
\end{table}
\noindent\textbf{Pesudo-Label Study.} As discussed in Section~\ref{sec:index}, the parameter $\text{neg}_{\text{start}}$ affects the difficulty level of pseudo-negative labels, which then impacts the effectiveness and efficiency of our index. To investigate the impacts of the pseudo-negative labels, we vary the hyperparameter $\text{neg}_{\text{start}}$ from 40,000, 50,000, 60,000, and 70,000, to 80,000 on the Beijing and Shanghai datasets. We illustrate the metrics $P(C)$ and $\text{IF}(C)$ of produced clusters in Figure~\ref{fig:negstart}. Notably, as $\text{neg}_{\text{start}}$ increases, both $IF(C)$ and $P(C)$ tend to increase. An increased $IF(C)$ suggests a more concentrated distribution of objects, while an increased $P(C)$ indicates improved accuracy in the retrieval results. The experiment results indicate that the choice of $\text{neg}_{\text{start}}$ leads to a trade-off between effectiveness and efficiency, which can be set flexibly in real-world applications.


\noindent\textbf{Spatial Learning Study.} 
To evaluate the learning-based spatial relevance module, 
we consider the following baseline: (1) {\Framework-R+$S_{in}$} that replaces the learning-based spatial relevance module with $S_{in}$ for training, where $S_{in}$ = 1 - $SDist(q.loc,o.loc)$; and (2) \Framework-R+$\alpha*S_{in}^{\beta}$ which substitutes the spatial relevance module with a learnable exponential function. $\alpha$ and $\beta$ are two learnable parameters and are processed to ensure non-negative. Table~\ref{tab:ablation} presents the experimental results obtained by conducting a brute-force search on the Beijing dataset using the trained models. Notably, \Framework-R outperforms \Framework-R+$S_{in}$ in all metrics. Interestingly, the first variant outperforms the second variant, which suggests that without careful design, a
learnable function may be outperformed by a simple approach.


\begin{table}[!htbp]
  \centering
  \caption{Ablation study of spatial relevance module and weight learning via brute-force search on the Beijing dataset.}
  \label{tab:ablation}
  \renewcommand{\arraystretch}{1.5} 
  \setlength{\arrayrulewidth}{0.5pt}
  \begin{tabularx}{0.5\textwidth}{>{\centering}m{2.5cm}*{4}{X}}
    \toprule[1pt]
    \multirow{2}{*}{} & \multicolumn{2}{c}{\textbf{Recall}} & \multicolumn{2}{c}{\textbf{NDCG}} \\
    \cmidrule(lr){2-3} \cmidrule(lr){4-5}
     & \textbf{@20} & \textbf{@10} & \textbf{@5} & \textbf{@1}\\
    \midrule[1pt]
    \Framework-R + $S_{in}$  & 0.7526 & 0.7087 & 0.5255 & 0.4271 \\
    \Framework-R + $\alpha *S_{in}^{\beta}$ &0.5308 &0.4532 &0.3130 &0.2411 \\
    \Framework-R + ADrW & 0.7925 & 0.7414 & 0.5832 & 0.4792 \\
    \midrule
    \textbf{\Framework-R} & \textbf{0.8156} & \textbf{0.7545} & \textbf{0.5913} & \textbf{0.4989} \\
    \bottomrule[1pt]
  \end{tabularx}
\end{table}
\noindent\textbf{Weight Learning Study.} We conduct a comparison experiment between our weight learning module and the attention mechanism proposed by DrW~\cite{liu2023effectiveness} (denoted as ADrW). Table~\ref{tab:ablation} reports the results obtained by conducting a brute-force search on the Beijing dataset using the trained relevance models. Here, \Framework-R represents our weight learning mechanism, while \Framework-R+ADrW denotes replacing it with the ADrW for training. The results indicate that our weight learning mechanism surpasses the latest ADrW mechanism.

\noindent{\textbf{Effect of Training Dataset Size on Effectiveness.} We investigate the impact of training dataset size over \Framework and \Framework-R. Specifically, we exclude a certain percentage of objects along with their corresponding records during the training and validation process. During testing, we use the complete test dataset. This approach avoids the issue of data leakage. We vary the percentage of objects from 30\%, 50\%, 80\% to 100\%. \revision{We present the results for NDCG@1 and Recall@10 on the Geo-Glue dataset using default hyperparameters, and similar results are observed for the remaining metrics and datasets.} The results are shown in Figure~\ref{fig:percentage_poi}. \Framework-R uses brute-force search with the relevance model and \Framework utilizes our proposed index for retrieval. The performance gap between them is consistently small, which underscores the capability of our index to boost retrieval speed without sacrificing effectiveness. Additionally, even with low training percentages (e.g., 0.3), our proposed method maintains satisfactory effectiveness, demonstrating its ability to learn from limited data and adapt to new queries.

\noindent\textbf{Effect of Training Dataset Size on Training Time.} We also measure the training time per epoch of \Framework-R and \Framework on \revision{the Geo-Glue dataset} by varying the size of the training data in the same manner stated in the last paragraph, and the results are shown in Figure~\ref{fig:training_time}. Other datasets exhibit similar trends. We observe that the training time of \Framework-R and \Framework appears to be linear to the size of the training data, aligning with our complexity analysis.

\begin{figure}[h]
\begin{center}
\includegraphics[width=0.45\columnwidth]{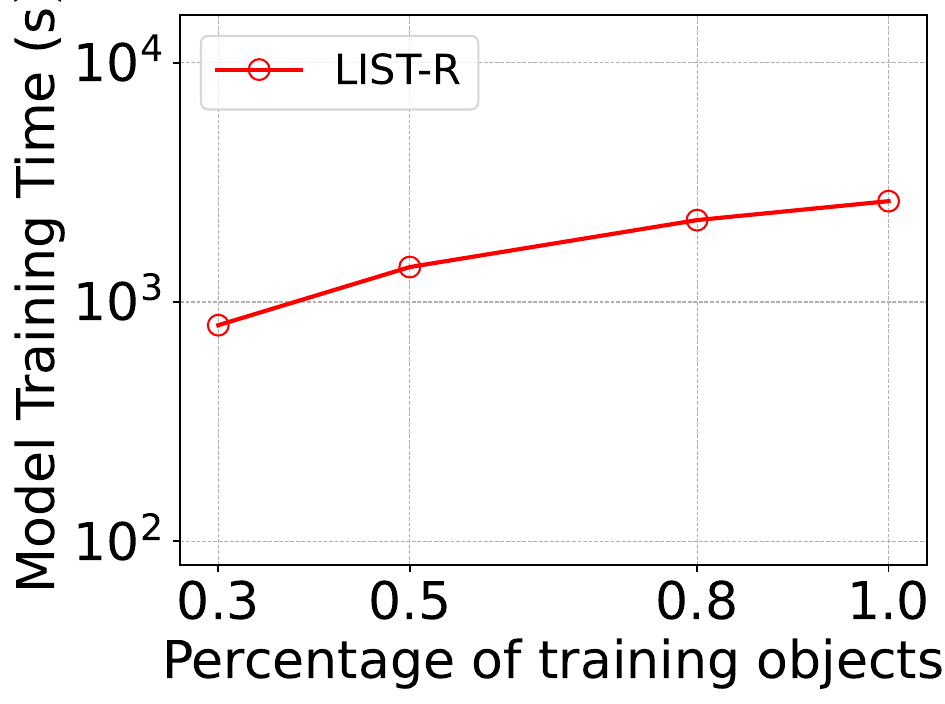}
\includegraphics[width=0.45\columnwidth]{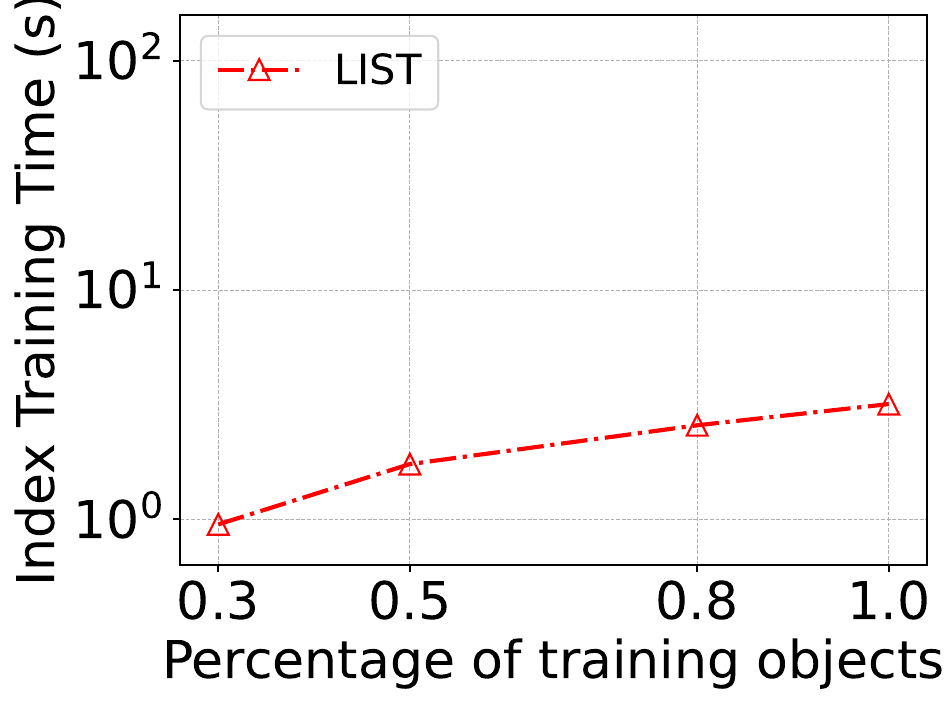}
\caption{The impact of training dataset size on training time for the Geo-Glue dataset.}
\label{fig:training_time}
\end{center}
\end{figure}

\begin{figure}[h]
\begin{center}
\includegraphics[width=0.45\columnwidth]{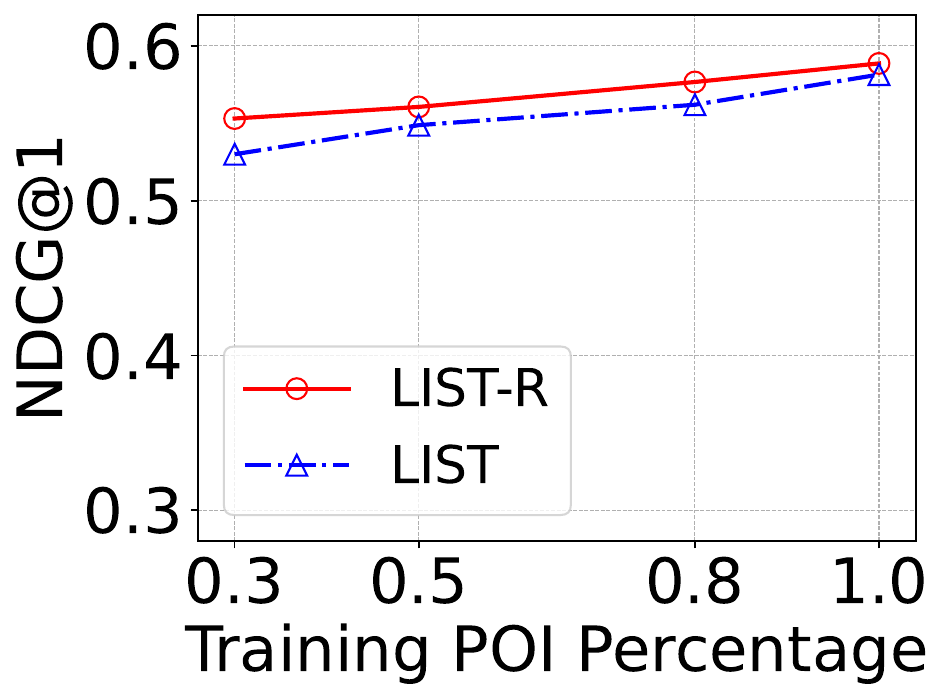}
\includegraphics[width=0.45\columnwidth]{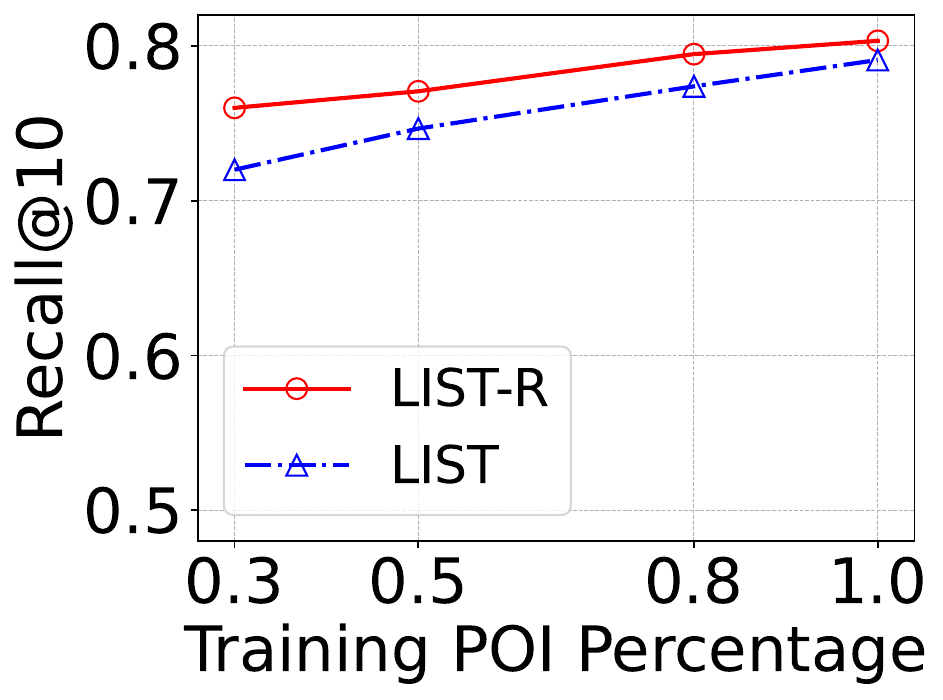}
\caption{\revision{The impact of training dataset size on effectiveness on the Geo-Glue dataset.}}
\label{fig:percentage_poi}
\end{center}
\end{figure}


\noindent{\textbf{\revision{Effect of Training Sample Selection on Effectiveness.}} \revision{To investigate the impact of training sample selection on effectiveness, we select different 50\% subsets from the Geo-Glue training dataset for training and then compare their effectiveness to demonstrate the influence of training sample choice. Since the random selection is determined by a random seed, we use different seeds (e.g., 1, 2, 3, 4) to ensure that each selection of training samples is different. The experimental results on the Geo-Glue dataset are presented in Figure~\ref{fig:train_sensitivity}, where we only report the results for NDCG@1, and the other metrics are similar. The results show that \Framework and \Framework-R produce similar results across different training samples, demonstrating that our solution is robust to variations in training sample selection.}

\begin{figure}[!htbp]
\centering
\begin{subfigure}[b]{0.45\columnwidth}
    \centering
    \includegraphics[width=\textwidth]{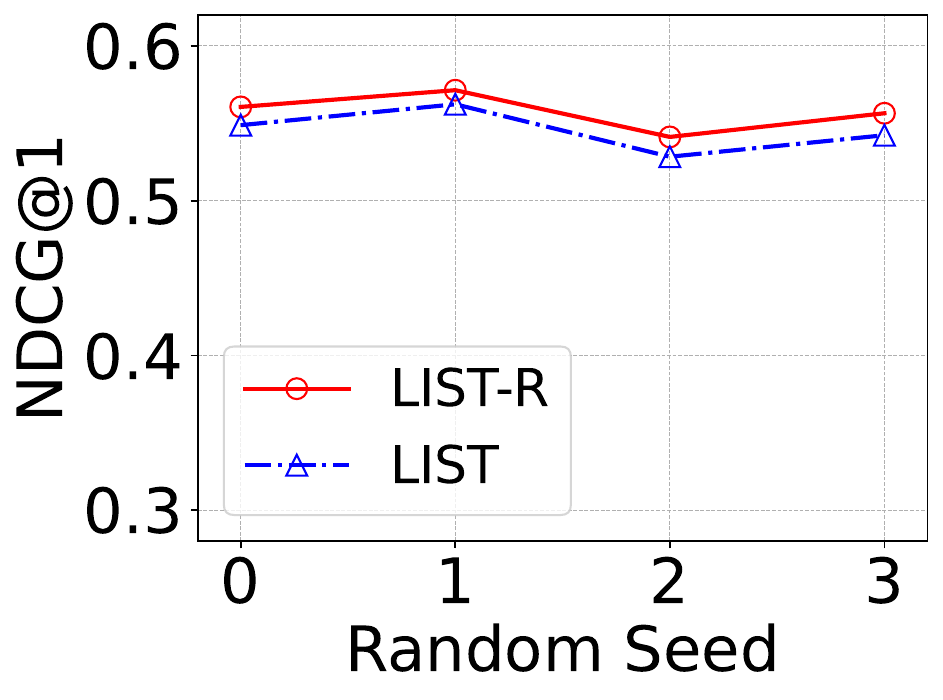}
    \caption{}
    \label{fig:train_sensitivity_ndcg}
\end{subfigure}
\hfill
\begin{subfigure}[b]{0.45\columnwidth}
    \centering
    \includegraphics[width=\textwidth]{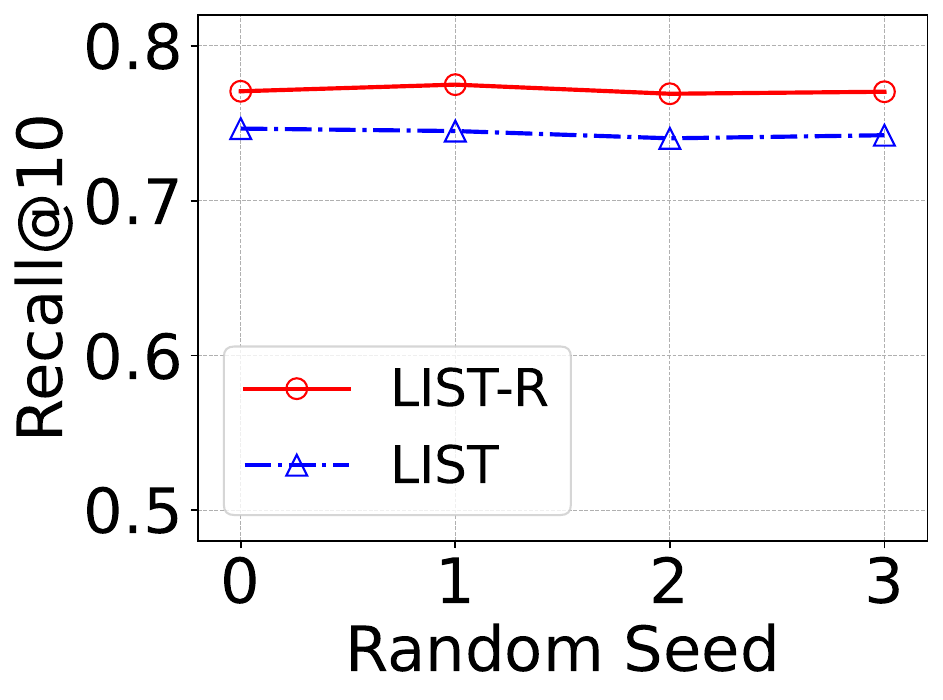}
    \caption{}
    \label{fig:train_sensitivity_recall}
\end{subfigure}
\caption{The impact of training sample selection on effectiveness for the Geo-Glue dataset.}
\label{fig:train_sensitivity}
\end{figure}

\begin{figure}[!htbp]
\centering
\begin{subfigure}[b]{0.45\columnwidth}
    \centering
    \includegraphics[width=\textwidth]{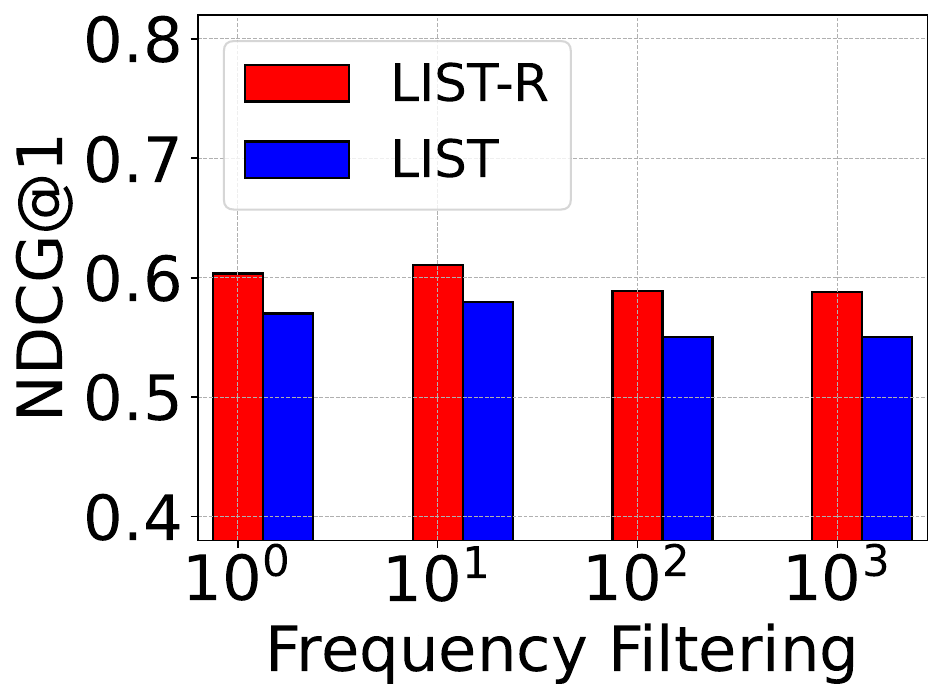}
    \caption{}
    \label{fig:low_frequency_ndcg}
\end{subfigure}
\hfill
\begin{subfigure}[b]{0.45\columnwidth}
    \centering
    \includegraphics[width=\textwidth]{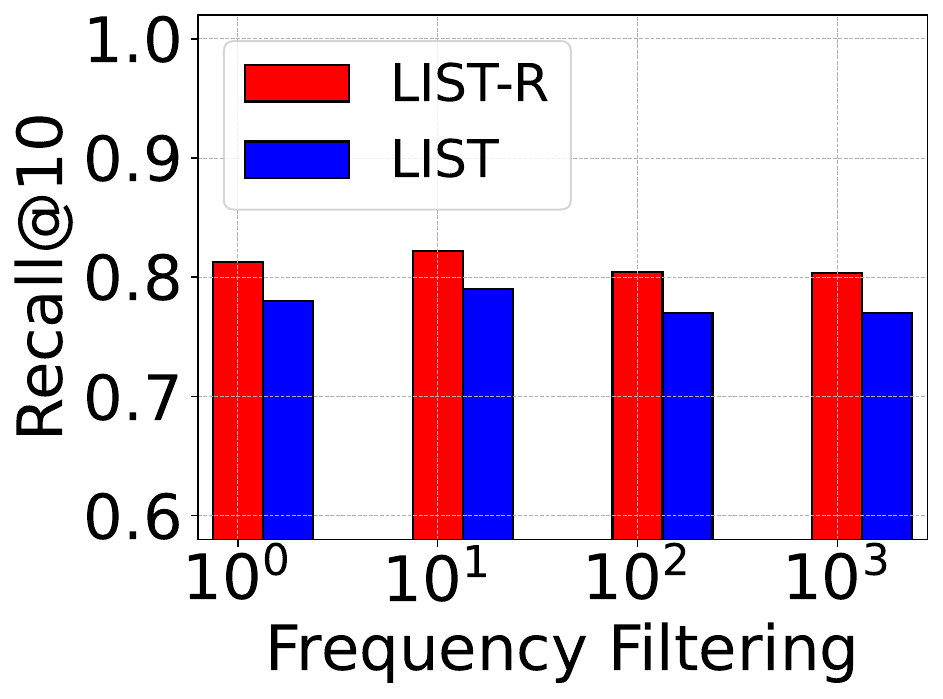}
    \caption{}
    \label{fig:low_frequency_recall}
\end{subfigure}
\caption{\revision{
Results of Low-Frequency Descriptors.}}
\label{fig:low_frequency_query}
\end{figure}
\noindent{\textbf{\revision{Evaluating Queries Containing Keywords of Low Frequency.}} \revision{To investigate the capability of \Framework and \Framework-R in handling unique or low-frequency keywords, such as 'halal' or 'vegan', which are more distinctive than common keywords like 'pizza' or 'pasta', we investigate whether the frequency of query keywords affects the performance of \Framework and \Framework-R on the geo-glue dataset. Specifically, we calculate the frequency of each keyword and identify queries containing low-frequency keywords, along with their corresponding ground-truth labels. The frequency of query keywords in the geo-glue dataset ranges from 1 to 3,891. 
We select queries containing keywords that appear at most $freq$ times, with $freq$ set to 1, 10, 100, and 1,000, resulting in subsets of test queries consisting of 4,108, 15,892, 19,930, and 19,991 queries, respectively. We then evaluate \Framework and \Framework-R on these subsets to demonstrate their capability in handling queries containing low-frequency keywords. The experimental results on the geo-glue dataset are shown in Figure~\ref{fig:low_frequency_query}, where we only report the results for Recall@10 and NDCG@1, as similar trends are observed on other metrics. The results show that \Framework and \Framework-R maintain performance for queries containing keywords of low-frequency. 
} 

\section{Conclusions and Future Work}
In this paper, \revision{we propose a novel index \Framework for embedding based spatial keyword queries. 
} We conduct extensive experiments to show the effectiveness and efficiency of \Framework over the state-of-the-art index baselines. This work opens up a promising research direction of designing novel ANNS indexes for accelerating the search for embedding based spatial keyword queries. One interesting future direction is integrating our index with product quantization techniques to further expedite the search process. Another potential direction is to extend our proposed index to vector databases for dense vectors without spatial information.

\bibliographystyle{spmpsci}
\bibliography{ref}

\begin{thebibliography}{10}
\providecommand{\url}[1]{{#1}}
\providecommand{\urlprefix}{URL }
\expandafter\ifx\csname urlstyle\endcsname\relax
  \providecommand{\doi}[1]{DOI~\discretionary{}{}{}#1}\else
  \providecommand{\doi}{DOI~\discretionary{}{}{}\begingroup \urlstyle{rm}\Url}\fi

\bibitem{DBLP:journals/pami/BabenkoL15}
Babenko, A., Lempitsky, V.S.: The inverted multi-index.
\newblock {IEEE} Trans. Pattern Anal. Mach. Intell. \textbf{37}(6), 1247--1260 (2015)

\bibitem{cary2010efficient}
Cary, A., Wolfson, O., Rishe, N.: Efficient and scalable method for processing top-k spatial boolean queries.
\newblock In: {SSDBM} 2010, vol. 6187, pp. 87--95 (2010)

\bibitem{DBLP:journals/geoinformatica/ChenXZZLFZ20}
Chen, X., Xu, J., Zhou, R., Zhao, P., Liu, C., Fang, J., Zhao, L.: S\({}^{\mbox{2}}\)r-tree: a pivot-based indexing structure for semantic-aware spatial keyword search.
\newblock GeoInformatica \textbf{24}(1), 3--25 (2020)

\bibitem{DBLP:journals/pvldb/ChenLCLBLGZ21}
Chen, Y., Li, X., Cong, G., Long, C., Bao, Z., Liu, S., Gu, W., Zhang, F.: Points-of-interest relationship inference with spatial-enriched graph neural networks.
\newblock Proceedings of the VLDB Endowment \textbf{15}(3), 504--512 (2021)

\bibitem{DBLP:conf/sigmod/ChenSM06}
Chen, Y., Suel, T., Markowetz, A.: Efficient query processing in geographic web search engines.
\newblock In: Proceedings of the {ACM} International Conference on Management of Data, SIGMOD 2006, pp. 277--288 (2006)

\bibitem{DBLP:journals/vldb/ChenCCJ21}
Chen, Z., Chen, L., Cong, G., Jensen, C.S.: Location- and keyword-based querying of geo-textual data: a survey.
\newblock {VLDB} J. \textbf{30}(4), 603--640 (2021)

\bibitem{congEfficientRetrievalTopk2009}
Cong, G., Jensen, C.S., Wu, D.: Efficient retrieval of the top-k most relevant spatial web objects.
\newblock Proceedings of the VLDB Endowment \textbf{2}(1), 337--348 (2009)

\bibitem{devlin2018bert}
Devlin, J., Chang, M., Lee, K., Toutanova, K.: {BERT:} pre-training of deep bidirectional transformers for language understanding.
\newblock In: {NAACL-HLT} 2019, Volume 1 (Long and Short Papers), pp. 4171--4186 (2019)

\bibitem{DBLP:conf/sigir/DingCXHLZX23}
Ding, R., Chen, B., Xie, P., Huang, F., Li, X., Zhang, Q., Xu, Y.: Mgeo: Multi-modal geographic language model pre-training.
\newblock In: H.H. Chen, W.J.E. Duh, H.H. Huang, M.P. Kato, J.~Mothe, B.~Poblete (eds.) SIGIR 2023, pp. 185--194 (2023)

\bibitem{dong2021continuous}
Dong, Y., Xiao, C., Chen, H., Yu, J.X., Takeoka, K., Oyamada, M., Kitagawa, H.: Continuous top-k spatial--keyword search on dynamic objects.
\newblock The VLDB Journal \textbf{30}(2), 141--161 (2021)

\bibitem{DBLP:conf/icde/FelipeHR08}
Felipe, I.D., Hristidis, V., Rishe, N.: Keyword search on spatial databases.
\newblock In: Proceedings of the 24th International Conference on Data Engineering, ICDE 2008, pp. 656--665 (2008)

\bibitem{gao2019beyond}
Gao, L., Zhu, X., Song, J., Zhao, Z., Shen, H.T.: Beyond product quantization: Deep progressive quantization for image retrieval.
\newblock In: Proceedings of the Twenty-Eighth International Joint Conference on Artificial Intelligence, {IJCAI} 2019, pp. 723--729 (2019)

\bibitem{DBLP:journals/pami/GeHK014}
Ge, T., He, K., Ke, Q., Sun, J.: Optimized product quantization.
\newblock IEEE Trans. Pattern Anal. Mach. Intell. \textbf{36}(4), 744--755 (2014)

\bibitem{DBLP:conf/cikm/GobelHNB09}
G{\"{o}}bel, R., Henrich, A., Niemann, R., Blank, D.: A hybrid index structure for geo-textual searches.
\newblock In: Proceedings of the 18th {ACM} Conference on Information and Knowledge Management, {CIKM} 2009, pp. 1625--1628 (2009)

\bibitem{DBLP:journals/tois/GuoCFSZC22}
Guo, J., Cai, Y., Fan, Y., Sun, F., Zhang, R., Cheng, X.: Semantic models for the first-stage retrieval: {A} comprehensive review.
\newblock {ACM} Trans. Inf. Syst. \textbf{40}(4), 66:1--66:42 (2022)

\bibitem{guo2020deep}
Guo, J., Fan, Y., Pang, L., Yang, L., Ai, Q., Zamani, H., Wu, C., Croft, W.B., Cheng, X.: A deep look into neural ranking models for information retrieval.
\newblock Information Processing \& Management \textbf{57}(6), 102,067 (2020)

\bibitem{guoAcceleratingLargeScaleInference2020}
Guo, R., Sun, P., Lindgren, E., Geng, Q., Simcha, D., Chern, F., Kumar, S.: Accelerating large-scale inference with anisotropic vector quantization.
\newblock In: Proceedings of the 37th International Conference on Machine Learning, ICML 2020, vol. 119, pp. 3887--3896 (2020)

\bibitem{DBLP:conf/www/HeLZNHC17}
He, X., Liao, L., Zhang, H., Nie, L., Hu, X., Chua, T.: Neural collaborative filtering.
\newblock In: Proceedings of the 26th International Conference on World Wide Web, {WWW} 2017, pp. 173--182 (2017)

\bibitem{Hsu18_L2C}
Hsu, Y., Lv, Z., Kira, Z.: Learning to cluster in order to transfer across domains and tasks.
\newblock In: 6th International Conference on Learning Representations, {ICLR} 2018 (2018)

\bibitem{Hsu19_MCL}
Hsu, Y., Lv, Z., Schlosser, J., Odom, P., Kira, Z.: Multi-class classification without multi-class labels.
\newblock In: 7th International Conference on Learning Representations, {ICLR} 2019 (2019)

\bibitem{Hsu16_KCL}
Hsu, Y.C., Kira, Z.: Neural network-based clustering using pairwise constraints.
\newblock arXiv preprint arXiv:1511.06321  (2015)

\bibitem{DBLP:conf/nips/HuLLC14}
Hu, B., Lu, Z., Li, H., Chen, Q.: Convolutional neural network architectures for matching natural language sentences.
\newblock In: Advances in Neural Information Processing Systems 27: Annual Conference on Neural Information Processing Systems 2014, pp. 2042--2050 (2014)

\bibitem{DBLP:conf/cikm/HuangHGDAH13}
Huang, P., He, X., Gao, J., Deng, L., Acero, A., Heck, L.P.: Learning deep structured semantic models for web search using clickthrough data.
\newblock In: 22nd {ACM} International Conference on Information and Knowledge Management, CIKM 2013, pp. 2333--2338 (2013)

\bibitem{huang2013learning}
Huang, P.S., He, X., Gao, J., Deng, L., Acero, A., Heck, L.: Learning deep structured semantic models for web search using clickthrough data.
\newblock In: Proceedings of the 22nd ACM International Conference on Conference on Information \& Knowledge Management, CIKM 2013, pp. 2333--2338 (2013)

\bibitem{indyk1998approximate}
Indyk, P., Motwani, R.: Approximate nearest neighbors: Towards removing the curse of dimensionality.
\newblock In: Proceedings of the Thirtieth Annual ACM Symposium on Theory of Computing - STOC '98, pp. 604--613 (1998)

\bibitem{jegouProductQuantizationNearest2011}
J{\'e}gou, H., Douze, M., Schmid, C.: Product quantization for nearest neighbor search.
\newblock IEEE Transactions on Pattern Analysis and Machine Intelligence \textbf{33}(1), 117--128 (2011)

\bibitem{DBLP:conf/kdd/Joachims02}
Joachims, T.: Optimizing search engines using clickthrough data.
\newblock In: Proceedings of the Eighth {ACM} International Conference on Knowledge Discovery and Data Mining, {SIGKDD} 2002, pp. 133--142 (2002)

\bibitem{johnson2019billion}
Johnson, J., Douze, M., J{\'e}gou, H.: Billion-scale similarity search with {GPUs}.
\newblock IEEE Transactions on Big Data \textbf{7}(3), 535--547 (2019)

\bibitem{DBLP:conf/emnlp/KarpukhinOMLWEC20}
Karpukhin, V., Oguz, B., Min, S., Lewis, P.S.H., Wu, L., Edunov, S., Chen, D., Yih, W.t.: Dense passage retrieval for open-domain question answering.
\newblock In: Proceedings of the 2020 Conference on Empirical Methods in Natural Language Processing, EMNLP 2020, pp. 6769--6781 (2020)

\bibitem{li2023geoglue}
Li, D., Ding, R., Zhang, Q., Li, Z., Chen, B., Xie, P., Xu, Y., Li, X., Guo, N., Huang, F., He, X.: Geoglue: {A} geographic language understanding evaluation benchmark.
\newblock CoRR \textbf{abs/2305.06545} (2023)

\bibitem{DBLP:journals/tkde/LiLZLLW11}
Li, Z., Lee, K.C.K., Zheng, B., Lee, W., Lee, D.L., Wang, X.: Ir-tree: An efficient index for geographic document search.
\newblock {IEEE} Trans. Knowl. Data Eng. \textbf{23}(4), 585--599 (2011)

\bibitem{lin2021inbatch}
Lin, S.C., Yang, J.H., Lin, J.: In-batch negatives for knowledge distillation with tightly-coupled teachers for dense retrieval.
\newblock In: Proceedings of the 6th Workshop on Representation Learning for NLP (RepL4NLP-2021), pp. 163--173 (2021)

\bibitem{liu2023effectiveness}
Liu, S., Cong, G., Feng, K., Gu, W., Zhang, F.: Effectiveness perspectives and a deep relevance model for spatial keyword queries.
\newblock Proceedings of the ACM International Conference on Management of Data, SIGMOD 2023 \textbf{1}(1), 1--25 (2023)

\bibitem{liu2019lsh}
Liu, W., Wang, H., Zhang, Y., Wang, W., Qin, L.: I-lsh: I/o efficient c-approximate nearest neighbor search in high-dimensional space.
\newblock In: 2019 IEEE 35th International Conference on Data Engineering (ICDE), pp. 1670--1673. IEEE (2019)

\bibitem{liu2014sk}
Liu, Y., Cui, J., Huang, Z., Li, H., Shen, H.T.: Sk-lsh: an efficient index structure for approximate nearest neighbor search.
\newblock Proceedings of the VLDB Endowment \textbf{7}(9), 745--756 (2014)

\bibitem{DBLP:conf/gis/Liu022}
Liu, Y., Magdy, A.: {U-ASK:} a unified architecture for knn spatial-keyword queries supporting negative keyword predicates.
\newblock In: Proceedings of the 30th International Conference on Advances in Geographic Information Systems, {SIGSPATIAL} 2022, pp. 40:1--40:11 (2022)

\bibitem{DBLP:conf/sigmod/LuLC11}
Lu, J., Lu, Y., Cong, G.: Reverse spatial and textual k nearest neighbor search.
\newblock In: Proceedings of the {ACM} {SIGMOD} International Conference on Management of Data, {SIGMOD} 2011, pp. 349--360 (2011)

\bibitem{luoSurveyDeepHashing2023}
Luo, X., Wang, H., Wu, D., Chen, C., Deng, M., Huang, J., Hua, X.S.: A survey on deep hashing methods.
\newblock ACM Transactions on Knowledge Discovery from Data \textbf{17}(1), 1--50 (2023)

\bibitem{malkovEfficientRobustApproximate2020}
Malkov, Y.A., Yashunin, D.A.: Efficient and robust approximate nearest neighbor search using hierarchical navigable small world graphs.
\newblock IEEE Transactions on Pattern Analysis and Machine Intelligence \textbf{42}(4), 824--836 (2020)

\bibitem{mikolov2013efficient}
Mikolov, T.: Efficient estimation of word representations in vector space.
\newblock arXiv preprint arXiv:1301.3781  (2013)

\bibitem{DBLP:conf/aaai/PangLGXWC16}
Pang, L., Lan, Y., Guo, J., Xu, J., Wan, S., Cheng, X.: Text matching as image recognition.
\newblock In: Proceedings of the Thirtieth AAAI Conference on Artificial Intelligence, 2016, pp. 2793--2799 (2016)

\bibitem{DBLP:journals/www/QianXZZZ18}
Qian, Z., Xu, J., Zheng, K., Zhao, P., Zhou, X.: Semantic-aware top-k spatial keyword queries.
\newblock World Wide Web \textbf{21}(3), 573--594 (2018)

\bibitem{qin2020similarity}
Qin, J., Wang, W., Xiao, C., Zhang, Y., Wang, Y.: High-dimensional similarity query processing for data science.
\newblock In: {KDD} '21: The 27th {ACM} Conference on Knowledge Discovery and Data Mining, {SIGKDD} 2021, pp. 4062--4063 (2021)

\bibitem{DBLP:conf/naacl/QuDLLRZDWW21}
Qu, Y., Ding, Y., Liu, J., Liu, K., Ren, R., Zhao, W.X., Dong, D., Wu, H., Wang, H.: Rocketqa: An optimized training approach to dense passage retrieval for open-domain question answering.
\newblock In: NAACL-HLT 2021, pp. 5835--5847 (2021)

\bibitem{renRocketQAv2JointTraining2021}
Ren, R., Qu, Y., Liu, J., Zhao, W.X., She, Q., Wu, H., Wang, H., Wen, J.: Rocketqav2: {A} joint training method for dense passage retrieval and passage re-ranking.
\newblock In: Proceedings of the 2021 Conference on Empirical Methods in Natural Language Processing, {EMNLP} 2021, pp. 2825--2835 (2021)

\bibitem{DBLP:journals/ftir/RobertsonZ09}
Robertson, S.E., Zaragoza, H.: The probabilistic relevance framework: {BM25} and beyond.
\newblock Found. Trends Inf. Retr. \textbf{3}(4), 333--389 (2009)

\bibitem{DBLP:conf/ssd/RochaGJN11}
Rocha{-}Junior, J.B., Gkorgkas, O., Jonassen, S., N{\o}rv{\aa}g, K.: Efficient processing of top-k spatial keyword queries.
\newblock In: Advances in Spatial and Temporal Databases - 12th International Symposium, {SSTD} 2011, vol. 6849, pp. 205--222 (2011)

\bibitem{DBLP:journals/pacmmod/Sheng0F00C023}
Sheng, Y., Cao, X., Fang, Y., Zhao, K., Qi, J., Cong, G., Zhang, W.: {WISK:} {A} workload-aware learned index for spatial keyword queries.
\newblock Proc. {ACM} Manag. Data \textbf{1}(2), 187:1--187:27 (2023)

\bibitem{tao2013fast}
Tao, Y., Sheng, C.: Fast nearest neighbor search with keywords.
\newblock {IEEE} Trans. Knowl. Data Eng. \textbf{26}(4), 878--888 (2014)

\bibitem{vaid2005spatio}
Vaid, S., Jones, C.B., Joho, H., Sanderson, M.: Spatio-textual indexing for geographical search on the web.
\newblock In: Advances in Spatial and Temporal Databases, 9th International Symposium, {SSTD} 2005, vol. 3633, pp. 218--235 (2005)

\bibitem{DBLP:journals/corr/WangSSJ14}
Wang, J., Shen, H.T., Song, J., Ji, J.: Hashing for similarity search: {A} survey.
\newblock CoRR \textbf{abs/1408.2927} (2014)

\bibitem{DBLP:journals/pami/WangZSSS18}
Wang, J., Zhang, T., Song, J., Sebe, N., Shen, H.T.: A survey on learning to hash.
\newblock {IEEE} Trans. Pattern Anal. Mach. Intell. \textbf{40}(4), 769--790 (2018)

\bibitem{DBLP:journals/pvldb/WangXY021}
Wang, M., Xu, X., Yue, Q., Wang, Y.: A comprehensive survey and experimental comparison of graph-based approximate nearest neighbor search.
\newblock Proceedings of the VLDB Endowment \textbf{14}(11), 1964--1978 (2021)

\bibitem{wang2020deltapq}
Wang, R., Deng, D.: Deltapq: lossless product quantization code compression for high dimensional similarity search.
\newblock Proceedings of the VLDB Endowment \textbf{13}(13), 3603--3616 (2020)

\bibitem{SIGIR_23_wen2023offline_dense_retrieval}
Wen, X., Chen, X., Chen, X., He, B., Sun, L.: Offline pseudo relevance feedback for efficient and effective single-pass dense retrieval.
\newblock In: Proceedings of the 46th International ACM SIGIR Conference on Research and Development in Information Retrieval, pp. 2209--2214 (2023)

\bibitem{wolf2019huggingface}
Wolf, T., Debut, L., Sanh, V., Chaumond, J., Delangue, C., Moi, A., Cistac, P., Rault, T., Louf, R., Funtowicz, M., Brew, J.: Huggingface's transformers: State-of-the-art natural language processing.
\newblock CoRR \textbf{abs/1910.03771} (2019)

\bibitem{DBLP:journals/tois/WuLWK08}
Wu, H.C., Luk, R.W.P., Wong, K., Kwok, K.: Interpreting {TF-IDF} term weights as making relevance decisions.
\newblock {ACM} Trans. Inf. Syst. \textbf{26}(3), 13:1--13:37 (2008)

\bibitem{DBLP:conf/www/YaoTCYXD021}
Yao, S., Tan, J., Chen, X., Yang, K., Xiao, R., Deng, H., Wan, X.: Learning a product relevance model from click-through data in e-commerce.
\newblock In: Proceedings of the Web Conference 2021, pp. 2890--2899 (2021)

\bibitem{DBLP:conf/sigir/YuanLLZYZX20}
Yuan, Z., Liu, H., Liu, Y., Zhang, D., Yi, F., Zhu, N., Xiong, H.: Spatio-temporal dual graph attention network for query-poi matching.
\newblock In: Proceedings of the 43rd International {ACM} {SIGIR} conference on research and development in Information Retrieval, {SIGIR} 2020, pp. 629--638 (2020)

\bibitem{DBLP:conf/icde/ZhangZZL13}
Zhang, C., Zhang, Y., Zhang, W., Lin, X.: Inverted linear quadtree: Efficient top k spatial keyword search.
\newblock In: 29th {IEEE} International Conference on Data Engineering, {ICDE} 2013, pp. 901--912 (2013)

\bibitem{DBLP:conf/icde/ZhangCMTK09}
Zhang, D., Chee, Y.M., Mondal, A., Tung, A.K.H., Kitsuregawa, M.: Keyword search in spatial databases: Towards searching by document.
\newblock In: Proceedings of the 25th International Conference on Data Engineering, {ICDE} 2009, pp. 688--699 (2009)

\bibitem{DBLP:conf/icde/ZhangOT10}
Zhang, D., Ooi, B.C., Tung, A.K.H.: Locating mapped resources in web 2.0.
\newblock In: Proceedings of the 26th International Conference on Data Engineering, {ICDE} 2010, pp. 521--532 (2010)

\bibitem{DBLP:conf/edbt/ZhangTT13}
Zhang, D., Tan, K., Tung, A.K.H.: Scalable top-k spatial keyword search.
\newblock In: Joint 2013 {EDBT/ICDT} Conferences, {EDBT} '13 Proceedings, pp. 359--370 (2013)

\bibitem{DBLP:conf/aaai/ZhaoPWCYZMCYQ19}
Zhao, J., Peng, D., Wu, C., Chen, H., Yu, M., Zheng, W., Ma, L., Chai, H., Ye, J., Qie, X.: Incorporating semantic similarity with geographic correlation for query-poi relevance learning.
\newblock In: The Thirty-Third {AAAI} Conference on Artificial Intelligence, AAAI 2019, pp. 1270--1277 (2019)

\bibitem{DBLP:journals/corr/abs-2211-14876}
Zhao, W.X., Liu, J., Ren, R., Wen, J.: Dense text retrieval based on pretrained language models: {A} survey.
\newblock CoRR \textbf{abs/2211.14876} (2022)

\bibitem{zhao2024dense}
Zhao, W.X., Liu, J., Ren, R., Wen, J.R.: Dense text retrieval based on pretrained language models: A survey.
\newblock ACM Transactions on Information Systems \textbf{42}(4), 1--60 (2024)

\bibitem{zheng2020pm}
Zheng, B., Xi, Z., Weng, L., Hung, N.Q.V., Liu, H., Jensen, C.S.: Pm-lsh: A fast and accurate lsh framework for high-dimensional approximate nn search.
\newblock Proceedings of the VLDB Endowment \textbf{13}(5), 643--655 (2020)

\bibitem{ECML_23_zhou2023master_dense_retrieval}
Zhou, K., Liu, X., Gong, Y., Zhao, W.X., Jiang, D., Duan, N., Wen, J.: {MASTER:} multi-task pre-trained bottlenecked masked autoencoders are better dense retrievers.
\newblock In: Machine Learning and Knowledge Discovery in Databases: Research Track - European Conference, {ECML} {PKDD} 2023, vol. 14170, pp. 630--647 (2023)

\bibitem{DBLP:conf/cikm/ZhouXWGM05}
Zhou, Y., Xie, X., Wang, C., Gong, Y., Ma, W.: Hybrid index structures for location-based web search.
\newblock In: Proceedings of the 2005 {ACM} {CIKM} International Conference on Information and Knowledge Management, pp. 155--162 (2005)

\end{thebibliography}

\end{document}